%% file: RIS.tex
\documentclass[onecolumn,draftclsnofoot,12pt]{IEEEtran}
\input{input.tex}

\usepackage{epstopdf}
\usepackage{enumerate}
\usepackage{algorithmicx}
\usepackage{algorithm}
\usepackage{amsmath}
\usepackage[noend]{algpseudocode}
\usepackage{float}
\usepackage{hyperref}
\usepackage{color}
\usepackage{makeidx}
\usepackage{bbm}
\usepackage{graphicx}
\usepackage{tablefootnote}
\usepackage{multirow}
\usepackage{multicol}

\newcommand{\sref}[1]{{Section}~\ref{#1}}

\begin{document}
\title{Design and Evaluation of Reconfigurable \\ Intelligent  Surfaces in Real-World Environment}
\author{Georgios C. Trichopoulos, Panagiotis Theofanopoulos, Bharath Kashyap, \\ Aditya Shekhawat, Anuj Modi, Tawfik Osman, Sanjay Kumar, Anand Sengar, \\ Arkajyoti Chang, and Ahmed Alkhateeb\\  \thanks{The authors are with the School or Electrical, Computer and Energy Engineering, Arizona State University, (Email: gtrichop, panagiotis.theofanopoulos, bgkashya, aditya.shekhawat, aymodi1, tmosman, svijay25, asengar2, arkajyoti.chang, alkhateeb@asu.edu).}}
\maketitle

\begin{abstract}
Reconfigurable intelligent surfaces (RISs) have promising  coverage and data rate gains for wireless communication systems in 5G and beyond. Prior work has mainly focused on analyzing the performance of these surfaces using computer simulations or lab-level prototypes. To draw accurate insights about the actual performance of these systems, this paper develops an RIS proof-of-concept prototype and extensively evaluates its potential gains in the field and under realistic wireless communication settings. In particular, a 160-element reconfigurable surface, operating at a 5.8GHz band, is first designed, fabricated, and accurately measured in the anechoic chamber. This surface is then integrated into a wireless communication system and the beamforming gains, pathloss, and coverage improvements are evaluated in realistic outdoor communication scenarios. When both the transmitter and receiver employ directional antennas and with 5m and 10m distances between the transmitter-RIS and RIS-receiver, the developed RIS achieves $15$-$20$dB gain in the signal-to-noise ratio (SNR) in a range of $\pm60^\circ$ beamforming angles. In terms of coverage, and considering  a far-field experiment with a  blockage between a base station and a grid of mobile  users and with an average distance of $35m$ between base station (BS) and the user (through the RIS), the RIS provides an average SNR improvement of $6$dB (max $8$dB) within an area $> 75$m$^2$. Thanks to the scalable RIS design, these SNR gains can be directly increased with larger RIS areas. For example, a 1,600-element RIS with the same design is expected to provide around $26$dB SNR gain for a similar deployment. These results, among others, draw useful insights into the design and performance of RIS systems and provide an important proof for their potential gains in real-world far-field wireless communication environments. 
\end{abstract}

\clearpage
\section{Introduction} \label{sec:Intro}

Reconfigurable Intelligent Surfaces (RISs) have attracted significant interest in the recent years from both academia and industry \cite{Liaskos2018,Basar2019,DiRenzo2020,Taha2021}. This is motivated by the promising gains that RISs are envisioned to offer for both millimeter wave (mmWave)/terahertz (THz) and sub-6GHz wireless communication systems. At mmWave/THz bands, RIS surfaces provide a potential solution for the critical coverage challenge by intelligently reflecting the wireless signals to the receiver direction \cite{ying2020relay,Nemati2020,moro2021planning,He_RIS}. At sub-6GHz bands, RIS systems could be leveraged to enhance the propagation characteristics \cite{Choi_RIS,Ozdogan}  and increase the spatial multiplexing gains. RISs also find interesting applications such as security \cite{Yang_RIS,Ai_RIS} and sensing \cite{RIS_radar,RIS_localization}. For example, RISs can suppress signals propagating toward eavesdroppers by creating "quite" zones around suspicious UEs. Thus, RISs can modify the channel appropriately and provide physical layer security in wireless networks  \cite{makarfi_physical_2020, wijewardena_physical_2021, khoshafa_reconfigurable_2021}. With all this potential, it is important to accurately evaluate the performance of the RIS surfaces in reality.  Based on this motivation, this work considers designing a low-power and portable proof of concept prototype for RIS-integrated wireless communications systems and leveraging it to validate the potential RIS gains in realistic communication environments. Next, we provide a brief background for the RIS circuits and systems before reviewing the relevant prior work in \sref{ssec:prior} and summarizing the key contributions in \sref{ssec:cont}. 

\subsection{A Brief Background} 
RISs comprise reconfigurable reflective surfaces that employ tunable subwavelength structures (e.g. antennas) to modulate the phase and/or amplitude of reflected waves. A smooth flat surface (e.g. mirror) reflects signals in the specular direction (incident angle = angle of reflection) because of the predetermined constant phase delay induced as the wave traverses the surface. On the other hand, RISs are capable of anomalous reflection (angle of reflection $\neq$ incident angle) by artificially modulating the phase and/or amplitude of the reflected wave \cite{yu_light_2011}. Besides redirecting the signal to desired directions in the far-field, RISs can also focus the energy when the user (or BS) is in the radiating near-field of the RIS. The sub-wavelength unit cell receives the incident signal and re-radiates it back into free space with a different amplitude and/or phase. Such modulation can be achieved by tuning the electromagnetic properties of the unit cell. For example, a switch placed at the feed of a patch antenna can alter the path of the electric current (open/short termination) and modulate the amplitude and phase of the re-radiated field. More switches can allow multi-bit wavefront modulation with improved beam control and efficiency. Several topologies can be found in the literature using single or multiple active devices as well as single or multi-layered substrates \cite{hum_reconfigurable_2014}. In all approaches, a biasing circuit and control unit (e.g. microcontroller) is required to be integrated within the RIS to control the state of the tunable devices by varying the biasing voltage across the device terminals. RISs are considered two-dimensional structures because the lateral dimensions are multiple wavelengths and thickness only a fraction of a wavelength. Such geometrical properties could allow for seamless installation on building surfaces (indoors or outdoors) even on curved surfaces.

In mmWave and THz non-line-of-sight (NLoS) paths, if the user is not in the specular reflection direction, then communication relies on diffuse scattering \cite{ma_terahertz_2019} from the rough surfaces of the surroundings (e.g. walls, terrain). Unless the user is near the specular direction, signal strength is impacted drastically and enabling anomalous reflection can provide viable propagation paths and boost the strength of received signal. Using the bistatic radar equation, we can estimate the received power $P_{r}$ when an RIS is placed between the user equipment (UE) and the base station (BS), as illustrated in \figref{fig:wc_model}:

\begin{equation} \label{eq1}
	P_r=\frac{P_t G_{BS} G_{UE} \lambda^2 \sigma}{(4\pi)^3 R_i^2 R_d^2} 
\end{equation}

Where $P_t$ is the signal power transmitted by the BS, $G_{BS}$, $G_{UE}$ are the gains of the BS and UE antennas respectively,  $R_i$ and $R_d$ the distances between BS-RIS and RIS-UE. The radar cross section $\sigma $ of the RIS can be approximated as a rectangular, flat conductive surface with losses. As such, the monostatic RCS of the an electrically large RIS with area $A$, efficiency $\eta$, and wavelength $\lambda$ is: 

\begin{equation} \label{eq2}
	\sigma=\frac{4\pi \eta A^2} { {\lambda}^2} 
\end{equation}

In a bistatic scenario, similar to the deployments in wireless communications, the RIS is viewed with an angle $\theta_i$ from the BS and $\theta_d$ from the UE (as shown in \figref{fig:wc_model}), assuming BS and UE are on the same plane. Then, the bistatic RCS of the RIS can be approximated by:

\begin{equation} \label{eq2}
	\sigma= \frac{4\pi \eta cos\theta_{i}cos\theta_{d}A^2}{{\lambda}^2} 
\end{equation}

We notice here that the received power increases quadratically with the size of the RIS. Namely, a ten-fold increase in the area $A$ provides 20 dB stronger signal for the same propagation scenario. Similarly, using RISs in higher frequencies, the received power will increase inversely proportional to the square of the wavelength. Such signal improvement is also crucial when considering that higher frequency signals suffer more losses (e.g. free path loss, penetration loss). For example, a window glass or a brick wall can have attenuation that exceeds 25 dB and 91 dB respectively at 38 GHz \cite{rodriguez_analysis_2015}.

\subsection{Prior Work on Reconfigurable Intelligent Surfaces} \label{ssec:prior}

The design and analysis of RIS-integrated wireless communication systems have attracted significant interest in the last few years. From the signal processing perspective, RIS systems bring interesting challenges to the design of the large-dimensional passive beamforming matrices at the RIS surfaces. For example, \cite{Huang_RIS} investigates the design of low-complexity and energy-efficient RIS beamforming algorithms, \cite{Zhou_RIS} designs robust RIS beamforming solutions, and \cite{Ye_RIS,Zhu_RIS,Yue_RIS} develop techniques for the joint basestation-RIS beamforming design problem. One main critical challenge for the operation of these systems is the high training overhead associated with the channel acquisition, especially if these surfaces are nearly passive. To address this challenge, \cite{Taha2021} developed RIS architectures where sparse active elements can be leveraged to acquire some knowledge about the channel. This can also enable leveraging compressive sensing and deep learning approaches to efficiently estimate the RIS channels \cite{Taha2021}. From the network perspective, it is important to accurately analyze the coverage performance when RIS surfaces are adopted. Some initial studies into this direction have been provided in \cite{ying2020relay,Nemati2020,moro2021planning,He_RIS} for various network architectures. All this prior work in \cite{Huang_RIS,Zhou_RIS,Ye_RIS,Zhu_RIS,Yue_RIS,Taha2021,ying2020relay,Nemati2020,moro2021planning,He_RIS}, however, was limited to simulation data. To accurately evaluate the potential of the RIS surfaces, it is crucial to build proof-of-concept prototypes and assess the expected coverage and data rate gains of RIS-integrated systems in  real world wireless communication environments. 

From the circuits and prototyping perspectives, various tuning topologies have been reported in the microwave frequency region using PIN or varactor diodes \cite{yang_programmable_2016, cui_coding_2014,wan_field-programmable_2016}. These topologies are used as beamformers in wireless communications, imaging, and sensing. In \cite{arun_rfocus_nodate}, an implementation of a reconfigurable intelligent surface was presented where the authors showed an indoor coverage analysis using a 2.4 GHz 3,200-element RIS. The user and/or base station, however, were in the near field of surface. In  \cite{pei_ris-aided_2021}, the authors presented a sub-6 GHz metasurface-based RIS to enhance received signal in wireless communication systems. Although the work shows clear improvement in the received signal, the experiments are limited to topologies where (i) both the receiver and transmitter are coupled to directive antennas and (ii) the receiver is always in the near field of the RIS. Additionally, the RIS prototype uses varactor switches, unit cells smaller than $\lambda$/2, vias, and a multilayered PCB structure, which is a viable solution for microwave frequencies, but a non-scalable approach for mmWave and THz frequencies. \textbf{This highlights the need to develop more efficient prototyping approaches for RIS systems and accurately evaluate their performance in realistic wireless communication scenarios where both the transmitter and receiver are in the far-field of the RIS surface. } 

Reconfigurable surfaces (either mirror or lenses) have also been implemented in the mmWave bands using also PIN diodes \cite{kamoda_60-ghz_2011} and there are many research efforts for higher frequency topologies approaching the THz bands \cite{headland_terahertz_2017}. However, new switching topologies and materials are required (e.g. $VO_2$, graphene, liquid crystal) to overcome the limited performance of PIN and varactor diodes \cite{vitale_modulated_2017} and simplified unit cell layouts to enable practically manufacturability of mmWave and THz RISs \cite{kashyap_mitigating_2020, theofanopoulos_novel_2020,theofanopoulos_modeling_2020,theofanopoulos_high-yield_2019}. For the RIS to be attractive in wireless communications, the surface needs to be scalable to large areas with thousands of unit cells and switches, low-profile and conformal to fit in various irregular indoor or outdoor surfaces, consume low power, and be cost efficient in manufacturing, installation, and maintenance. Although most of these features are inherent in reflective surfaces, switching performance and manufacturability due to the biasing circuit topologies are engineering challenges. 

\subsection{Contribution} \label{ssec:cont}
In this paper, we designed and fabricated  a low-power reconfigurable intelligent surface and demonstrated its beamforming and coverage gain potential capabilities in real-world wireless communications environments. The main contributions can be summarized as follows:
\begin{itemize}
	\item Present a single layer, single switch-per-cell RIS design operating at 5.8 GHz that is also compatible with mmWave and THz fabrication and integration technologies. The layout requires no vertical connections (e.g. vias) and biasing network in integrated on the same plane. The RIS is capable of electronic beam scanning in both azimuth and elevation planes.
	\item Characterize the beamforming capabilities in an outdoor environment. We carried out radiation pattern characterization  in the controlled environment of an anechoic chamber and then evaluated the beamforming gains in realistic near-field and far-field outdoor settings in the presence of scattering from the ground and the surrounding environment.
	\item Carry out coverage measurements at 5.8 GHz in occluded (line-of-sight (LoS) obstructed), outdoor areas using mobile UE with omni-directional antennas. The LoS path between the BS and mobile UE is blocked by a building and the signal coverage is improved when using the proposed RIS to provide an alternative signal path. 
\end{itemize}

The paper is structured as follows: In section II, we present current RIS design and discuss the fundamental operation of reflectarray antennas, which is a main component in RISs. In section III, we discuss the quantized beamforming theory of RISs and its relationship to wireless communication between a BS and UE. Then, section IV presents the design and characterization of the RIS, including a discussion on the unit cell layout (building block of the RIS), testing of the integrated switches and beamforming, and integration with the necessary control circuitry. In section V, we present a wireless communication testbed - including a BS, UE, and the RIS - to evaluate the performance of the RIS in realsitic wireless communication field tests. Finally, in section V, we evaluate the potential gains of the RIS  beamforming in improving the SNR at the mobile users  and extending the wireless communication coverage beyond  LoS areas.

\section{RIS-Based Wireless Communication System}
In this section, we describe the adopted RIS-assisted wireless communication system and briefly formulate the beamforming design problem. 

\begin{figure}[t]
	\centering
	\includegraphics[width=1\columnwidth]{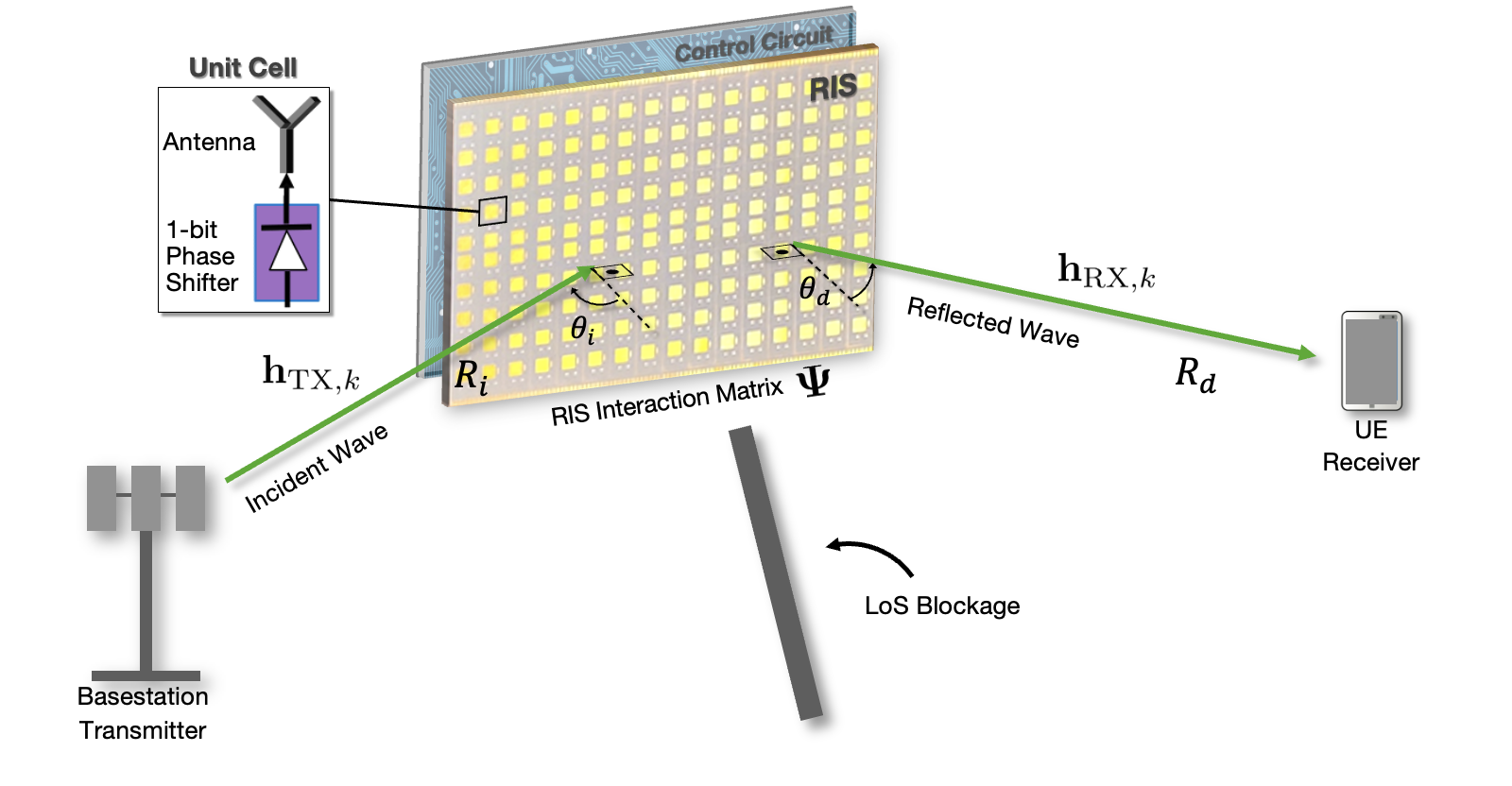}
	\caption{Overview of the proposed reconfigurable intelligent surface (RIS). A control circuit provides the necessary excitation (codebook) to the unit cells and redirects the incident wave to the desired direction. }
	\label{fig:wc_model}
\end{figure}

\subsection{System and Signal Models} \label{subsec:sys_model}

As depicted in \figref{fig:wc_model}, we consider a simple scenario where the communication between a single-antenna transmitter and a single-antenna receiver is assisted by  a reconfigurable intelligent surface. The surface is assumed to have a uniform planar array structure of $M \times N$ elements. Adopting an OFDM (Orthogonal Frequency Division Multiplexed) system model with $K$ sub-carriers, we define $\bh_{\mathrm{TX},k}$ and $\bh_{\mathrm{RX},k}$ as the $k$th subcarrier $M N \times 1$ uplink channels from the transmitter and receiver to the RIS, $k=1, 2, ..., K$. Further, we define $h_{\mathrm{TR}, k}$ as the direct LoS channel between the transmitter and receiver on the $k$th subcarrier. If $s_k$ denotes the transmitted signal over the $k$th subcarrier, then the received signal can be written as
\begin{equation} \label{eq:rec}
	r_k = \bh_{\mathrm{RX},k}^T \boldsymbol{\Psi} \bh_{\mathrm{TX},k} s_k + h_{\mathrm{TR},k} s_k + n_k,
\end{equation}
where $n_k \sim \mathcal{N}_\mathbb{C}(0,\sigma_n^2)$ represents the receive noise and $s_k$ has an average power $\bbE\left[s_k\right]=\frac{P}{K}$ with P representing the total transit power. The matrix $\boldsymbol{\Psi}$ denotes the $M N \times M N$ RIS interaction matrix. Note that $\boldsymbol{\Psi}$ is a diagonal matrix. To capture that, we define $\boldsymbol{\psi} = \mathrm{diag} \left(\boldsymbol{\Psi}\right)$ as the $M N \times 1$ RIS reflection vector, which includes the phase control at each RIS element. In particular, each element $mn$ is expressed as $e^{j \varphi_{mn}}$, with the $\varphi_{mn}$ denoting the RIS modulation phase of this element.  Now, neglecting the LoS link (for the scenarios with blocked LoS), the RIS beamforming/excitation vector $\boldsymbol{\psi}$ can be designed to maximize the achievable rate following 
\begin{align} 
	\boldsymbol{\psi}^\star = & \arg \hspace{-10pt}  \max_{\varphi_{mn}, \forall m,n} \ \ \ \ \sum_{k=1}^K \log_2\left(1+ \rho {\left|\bh_{\mathrm{RX},k}^H \boldsymbol{\Psi} \bh_{\mathrm{TX},k} \right|^2}\right), \label{eq:opt1-1} \\
	& \hspace{20pt} \text{s.t.} \hspace{40pt} \varphi_{mn} \in \left[0^\circ, 360^\circ\right], \label {eq:opt1-2}\ \ \forall m,n,
\end{align}
where $\rho=\frac{P}{K \sigma_n^2}$ denoting the per-carrier SNR. Next, since the RIS-transmitter and RIS-receiver channels are mostly dominated by LoS paths, we consider the following approaximation for the design of the RIS reflection vector 
\begin{align} 
	\boldsymbol{\psi}^\star = & \arg \hspace{-10pt}  \max_{\varphi_{mn}, \forall m,n} \ \ \ \ \sum_{k=1}^K {\left|\bh_{\mathrm{RX},k}^H \boldsymbol{\Psi} \bh_{\mathrm{TX},k} \right|^2}, \label{eq:opt2-1} \\
	& \hspace{20pt} \text{s.t.} \hspace{40pt} \varphi_{mn} \in \left[0^\circ, 360^\circ\right], \label {eq:opt2-2}\ \ \forall m,n.
\end{align}

In practice, the channels between the RIS and the transmitters/receivers will likely be LoS. Now, focusing on this case, the  optimization problem in \eqref{eq:opt2-1}-\eqref{eq:opt2-2} can be further reduced to 
\begin{align}
	\boldsymbol{\psi}^\star = & \arg \hspace{-10pt} \max_{\varphi_{mn}, \forall m,n} \ \ \ \  {\left|\left(\ba_{\mathrm{RIS} }^*(\theta_d, \phi_d)  \odot   \ba_{\mathrm{RIS}}(\theta_i, \phi_i) \right)^T \boldsymbol{\psi} \right|^2}, \label{eq:opt3-1} \\
	&  \hspace{20pt} \text{s.t.} \hspace{40pt} \varphi_{mn} \in \left[0^\circ, 360^\circ\right], \ \ \forall m,n.  \label{eq:opt3-2}
\end{align}
where $\ba_\mathrm{RIS}(\theta,\phi)$ denotes the RIS array response vector for the angles $\theta,\phi$. The angles $\theta_i$ and $\phi_i$ represent the  elevation/azimuth angles of the incident signal and the angles $\theta_d$ and $\phi_d$ represent the  elevation/azimuth angles of the desired reflection direction. In the next subsection, we briefly present the design approach for the beamforming codebook adopted in this work. 

\subsection{1-bit Beamforming Codebook} \label{subsec:codebook}
Consider the RIS beamforming design problem in \eqref {eq:opt3-1}-\eqref{eq:opt3-2}, if the incident and desired reflection directions are $(\theta_i, \phi_i)$ and $(\theta_d, \phi_d)$, respectively, then the optimal RIS phase shifting configuration for each element $\varphi_{mn}$ in $\boldsymbol{\psi}$ is given by 
\begin{equation} \label{eq:ct_phase}
	\varphi_{mn} = \varphi_{i, mn} - \varphi_{d, mn},
\end{equation}
where $ \varphi_{i,mn}$ and $\varphi_{d,mn}$ are respectively the phase of the incident wave and phase for the desired reflection direction on the $mn^{th}$ RIS unit cell. For a two-dimensional  planar RIS with unit cells arranged on the $x-y$ plane, these phases are given by \cite{Encinar}
\begin{equation} \label{eq2}
	\varphi_{i, mn} = k_0(x_m \sin\theta_i \cos\phi_i +y_n \sin\theta_i \sin\phi_i),
\end{equation}
\begin{equation} \label{eq3}
	\varphi_{d, mn} = k_0(x_m \sin\theta_d \cos\phi_d +y_n \sin\theta_d \sin\phi_d),
\end{equation}
with $k_0$ representing the free space wavenumber and $(x_m,y_m )$ are the coordinates of the $mn^{th}$ RIS element.
However, at mmWave/THz frequencies practical phase shifting topologies can only produce discrete values of $ \varphi_{mn}$ that are typically quantized using 1-, 2- or 3- bit quantization schemes. A single-bit phase quantization scheme is adopted in this work owing to its simplicity and lower cost when compared to higher bit quantization methods \cite{FanYang}. As such, all the phase values in the range $[-90^0, +90^0]$ are rounded off to $0^0$ (designated as state ‘0’/OFF) and the rest of the phase values are rounded off to $180^0$ (designated as state ‘1’/ON). The two states are realized using PIN diodes with state ‘0’ representing the OFF state of the diode and state ‘1’ corresponding to the diode’s ON state, as will be explained in detail in \sref{sec:design}. Thus, the quantized phase shift at the $mn^{th}$ RIS unit cell is :
\begin{equation} \label{eq4}
\varphi^\mathrm{quant}_{mn} = \left|180^0\cdot \mathrm{round}(\frac{\varphi_{mn}}{180^0})\right|.
\end{equation}

With this RIS beamforming design, the resulting far-field radiation pattern of the RIS at a direction $\theta, \phi$ can be approximated by the array factor
\begin{equation} \label{eq5}
AF_\mathrm{RIS}(\theta, \phi) = \sum_{m=1}^{M} \sum_{n=1}^{N} e^{-j k_0 (x_m u +y_n v)} e^{j \varphi_{i, mn}} e^{j \varphi^\mathrm{quant}_{mn}},
\end{equation}
where $k_0 (x_m u+y_n v)$ represents the phase modulation due to Green’s function, with $u=\sin\theta \cos\phi$ and $v=  \sin\theta \sin\phi$.

For efficient wireless communication operation with the RIS, we design a codebook $\boldsymbol{\mathcal{P}}$ of pre-defined beams (RIS phase shifting configuration). Each RIS beam codeword in this codebook reflects the wave that is incident from a direction $(\theta_i, \phi_i)$ to a desired reflection direction $(\theta_d, \phi_d)$. For example, if the desired sets of directions for the incident and reflected waves are respectively defined by 
\begin{align}
& \mathcal{I} = \left\{\left(\theta_{i,1}, \phi_{i,1}\right),\left(\theta_{i,2}, \phi_{i,2}\right), ..., \left(\theta_{i,N_i}, \phi_{i,N_{i}}. \right) \right\} \\
& \mathcal{D}=\left\{\left(\theta_{d,1}, \phi_{d,1}\right),\left(\theta_{d,2}, \phi_{d,2}\right), ..., \left(\theta_{d,N_d}, \phi_{d,N_{d}}\right) \right\}
\end{align}
with cardinalities $\left|\mathcal{I}\right|=N_i$ and $\left|\mathcal{D}\right|=N_d$, then the RIS codebook $\boldsymbol{\mathcal{P}}$ will have $N_i N_d$ beam codewords. Each codeword vector $\overline{\boldsymbol{\psi}}_{i,d} \in \boldsymbol{\mathcal{P}}$ is designed following the 1-bit (quantized) beamforming approach in \eqref{eq:ct_phase}-\eqref{eq4} for a particular incident/reflection directions from $\mathcal{I}$ and $\mathcal{D}$.

\section{Reconfigurable Intelligent Surface Design} \label{sec:design}

In this section, we present the design and characterization of the RIS, including the unit cell layout, measurement of the radiation pattern of both a fixed-beam (no switches) and a multi-beam RIS prototype (integrated PIN switches), and integration of the control circuitry. The RIS is designed at a center frequency of 5.8 GHz which is part of the unlicensed spectrum used in wireless local area networks (WLANs). The goal of this work is to design a single layer topology (besides the ground plane) without requiring the use of vertical components (e.g. vias). As such, the RIS is comprised of the ground plane, the substrate, and the top metalization layer. Such topology is attractive for future mmWave and THz RISs where multilayer structures increase losses and fabrication complexity.

\begin{figure}[t]
	\centering
	{\includegraphics[width = 0.9\linewidth]{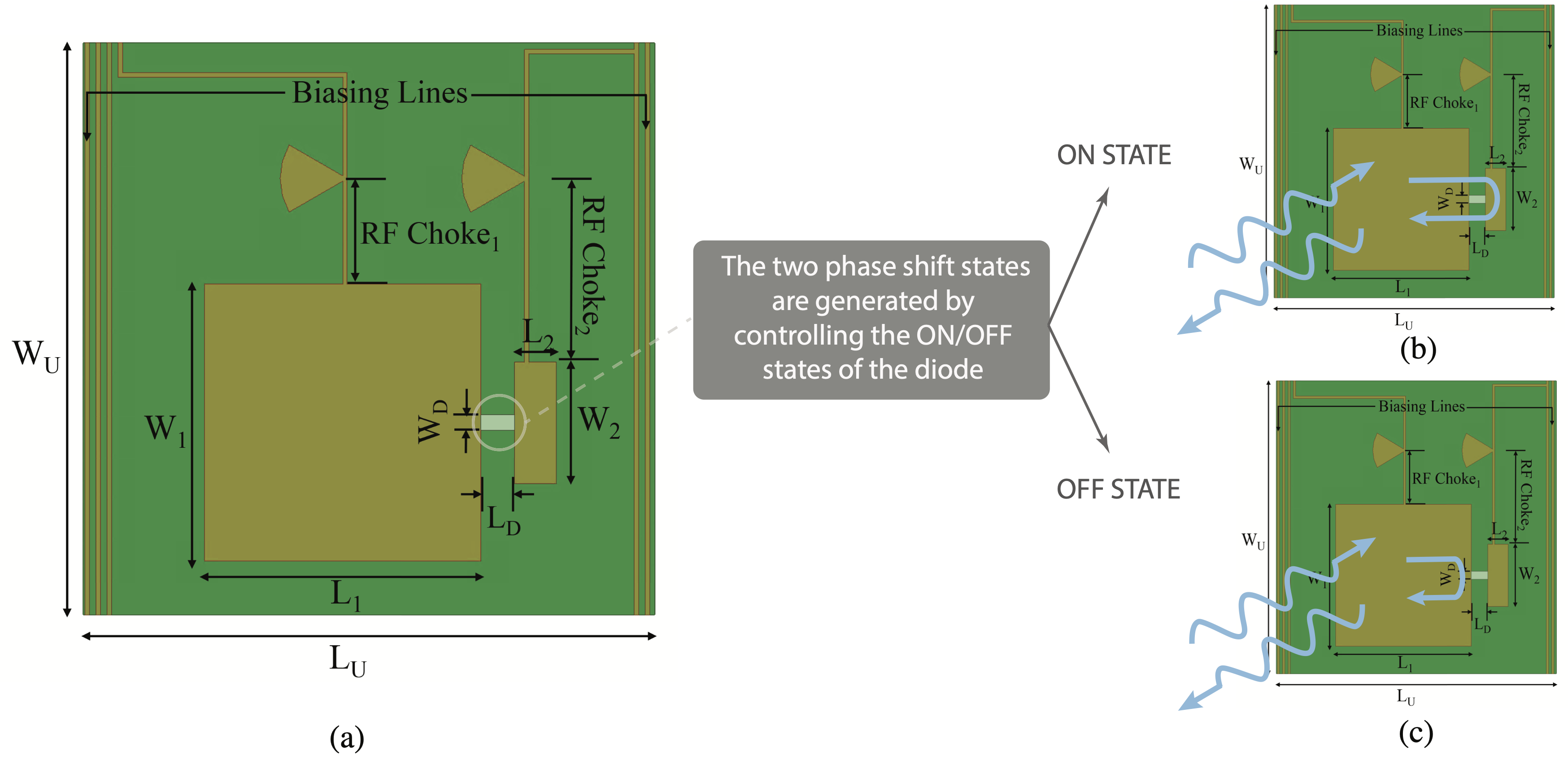}}
	
	\caption{Layout of the RIS unit cell. (a) The unit cells comprises a main resonant patch antenna connected to a parasitic rectangular patch through an RF PIN diode (switch). The necessary biasing lines and RF chokes are integrated for switch activation (ON/OFF). (b), (c) Activation of the RF PIN diode alters the current distribution, resulting in phase modulation of the reflected signal (180 degrees) without a significant modulation of the magnitude within the bandwidth of interest.} 
	\label{fig:states}
\end{figure}

\begin{table}[t]
	\caption{Key dimensions of the RIS unit cell}
	\label{table}
	\centering
	\setlength{\tabcolsep}{5pt}
	\renewcommand{\arraystretch}{1.2}
	\begin{tabular}{|c|c|}
		\hline
		\textbf{Unit Cell Design Parameters}& \textbf{Dimension (mm)} \\ \hline \hline
		$\mathrm{W_U}$ & 25.85 \\	\hline
		$\mathrm{W_1}$ & 12.51 \\ \hline
		$\mathrm{L_1}$ & 12.51 \\ \hline
		$\mathrm{W_2}$ & 5.48 \\ \hline
		$\mathrm{W_D}$ & 0.7 \\    \hline
		$\mathrm{L_D}$ & 1.5 \\ \hline
	\end{tabular}
	\label{tab_params}
\end{table}
\subsection{1-bit Unit Cell Design}
\begin{figure}
	\centering
	{\includegraphics[width = 0.7\linewidth]{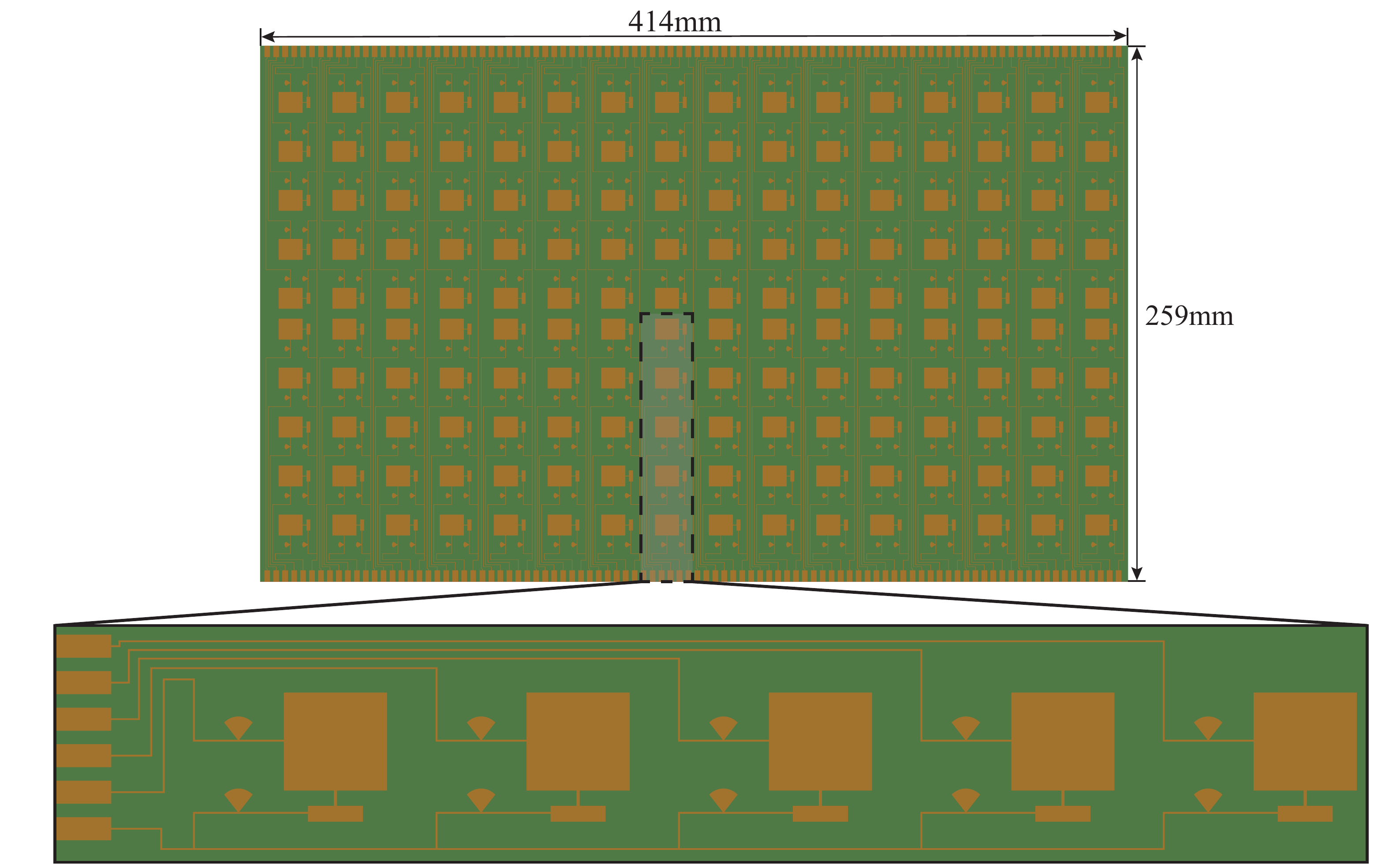}}
	
	\caption{The layout of the 160-element (16 $\times$ 10) antenna array used in the proposed RIS. The inset shows the routing of the biasing lines for 5 unit cells (no vias). This constitutes a 2D topology that is compatible with low complexity RIS implementation in the mmWave and THz bands.} 
	\label{fig:array}
\end{figure}

The RIS comprises 160 (16 $\times$ 10) unit cells that contain a passive antenna, a radio-frequency (RF) PIN diode (switch), and biasing lines, as shown in \figref{fig:states} and \figref{fig:array}. Depending on the applied voltage across the diode terminals (reverse bias: OFF, forward bias: ON), the antenna re-radiates the received signal with a phase difference that depends on the current distribution. Typically, such a phase shift is enabled either by changing the resonant frequency of the antenna \cite{pozar_design_1997, hum_modeling_2007} or by providing extra path to the current on the feed of the antenna \cite{venneri_design_2013, kashyap_mitigating_2020}. The latter approach requires a short transmission line terminated to the ground. Such topology would require a via, thus, to minimize fabrication complexity, we opted for the resonance approach by adding a parasitic patch next to the antenna connected through the PIN diode. Additionally, biasing lines are needed for both diode terminals, therefore, two narrow lines are connected to the antenna and the parasitic patch respectively, as depicted in \figref{fig:states}. 


To isolate the RF signal from the biasing lines, a radial stub is used in each one of the lines. The biasing lines are routed in groups of five unit cells to ensure minimum wiring complexity, as shown in the full array topology in \figref{fig:array}. As opposed to current approaches in the literature, the proposed topology 1) requires only a single tuning device (switch) and 2) comprises a single layer with no vertical connections to the ground. Although single switch approaches result in quantization sidelobes, in this study, we use the RIS only in limited scanning range to avoid sidelobe interference. Nevertheless, using pre-coded, phase randomization methods \cite{kashyap_mitigating_2020, yin_single-beam_2020} we can eliminate the undesired side lobes using practical topologies. This step has not been the focus of this work and is left for future prototype implementations.

\begin{figure}[t]
		\centering
		\begin{subfigure}[Reflection Phase Response]
		{\includegraphics[width=0.45\linewidth]{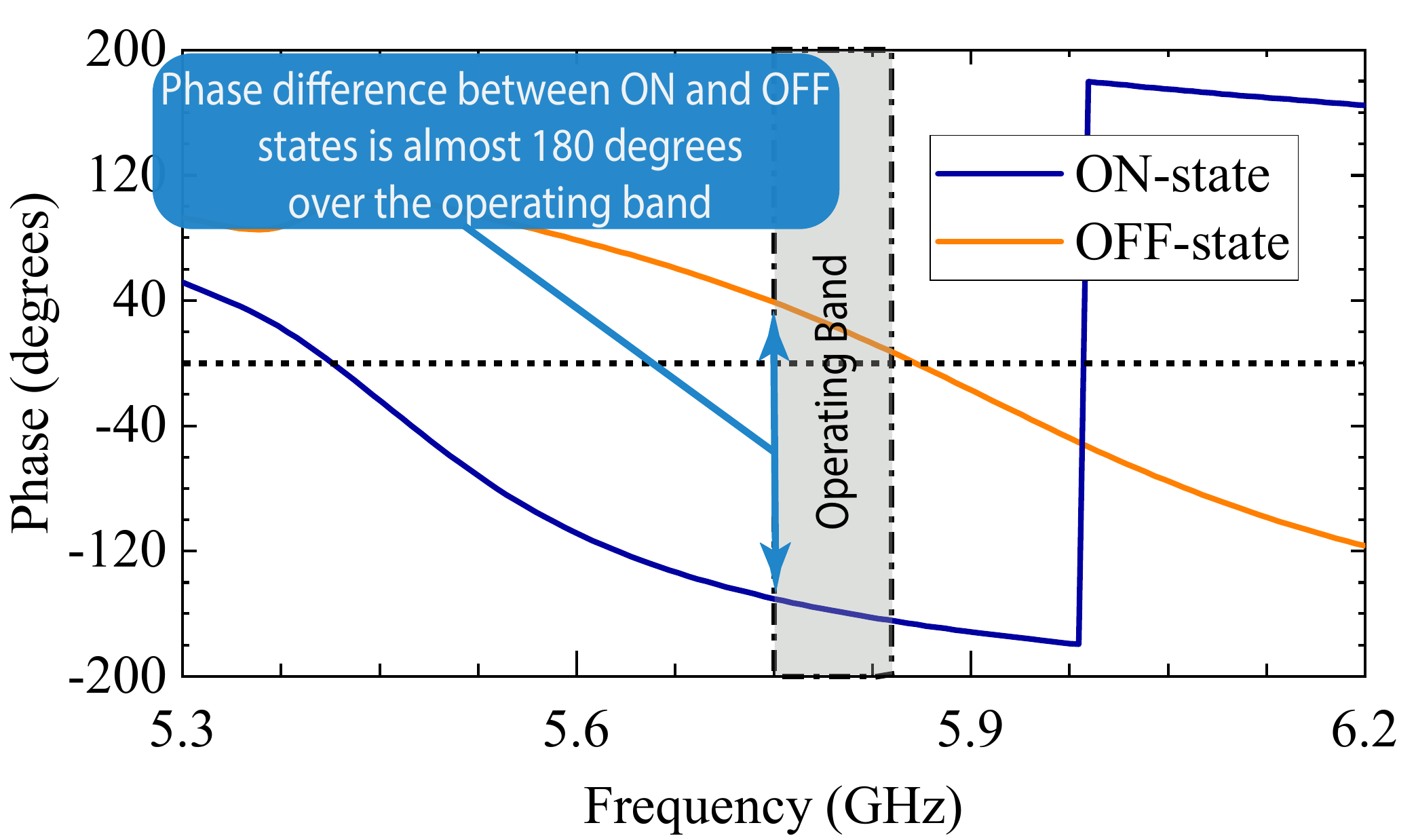}}
		\end{subfigure}
		\begin{subfigure}[Reflection Magnitude Response]
		{\includegraphics[width=.45\linewidth]{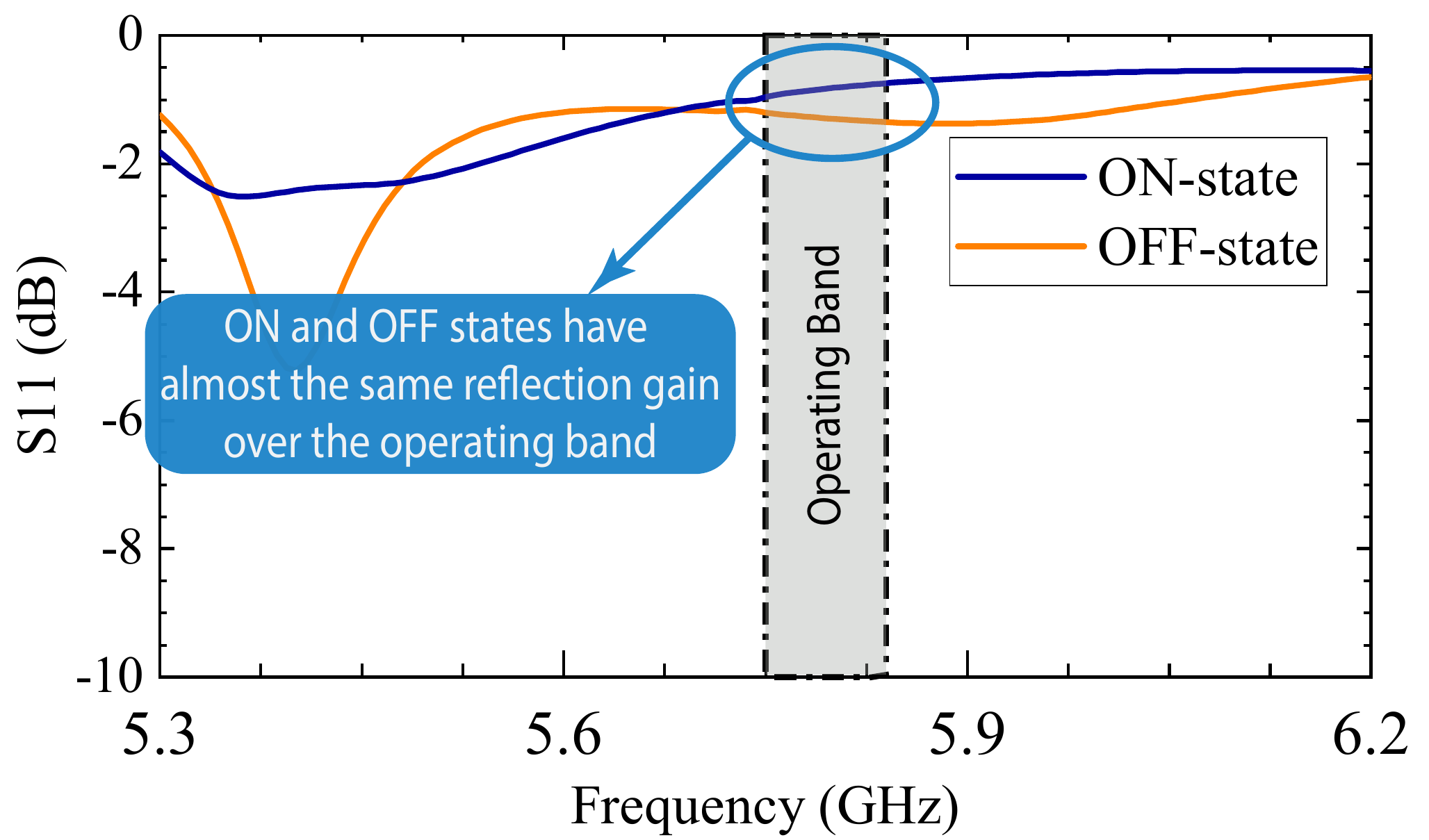}}
		\end{subfigure}
		\caption{Numerical analysis of the unit cell response under boresight illumination ($\vartheta_{i}$=0) for two switch states (ON/OFF).} 
		\label{fig:phase_mag_unit_cell}
	\end{figure}

To evaluate the response of the unit cell under different biasing states, we use an industry standard commercial electromagnetics (EM) solver (ANSYS, HFSS). As such, we design an infinite array of unit cells using periodic boundary conditions on each primary direction. Although this process omits radiation effects at the edge of the aperture, it is considered reliable in estimating the EM response of an individual unit cell. The substrate used here is Rogers RT/ Duroid 6002 RF laminate with dielectric constant, height and loss tangent of 2.94, 2.54mm, 0.0012 respectively. The antenna/switch (BAR50-02V) co-design is carried out using the measured $S$-parameters provided by the manufacturer (MACOM). When the array is illuminated from boresight, the calculated reflected signal exhibits small ($<1$ dB) magnitude modulation and a phase difference of 180$^{\circ}$ at 5.8 GHz for the two states of the switch. For an acceptable phase modulation range of 180$^{\circ}$ $\pm$ 20 $^{\circ}$ between the two states, the expected bandwidth is approximately 150 MHz, as shown in \figref{fig:phase_mag_unit_cell}. Next, we integrate the unit cells into a large array and analyze the radiation effects under various switch excitations (codebook).

\subsection{Array Design}
\begin{figure}[t]
	\centering
	\begin{subfigure}[At a Reflection Angle of $0^{\circ}$]
		{\includegraphics[width=0.31\textwidth]{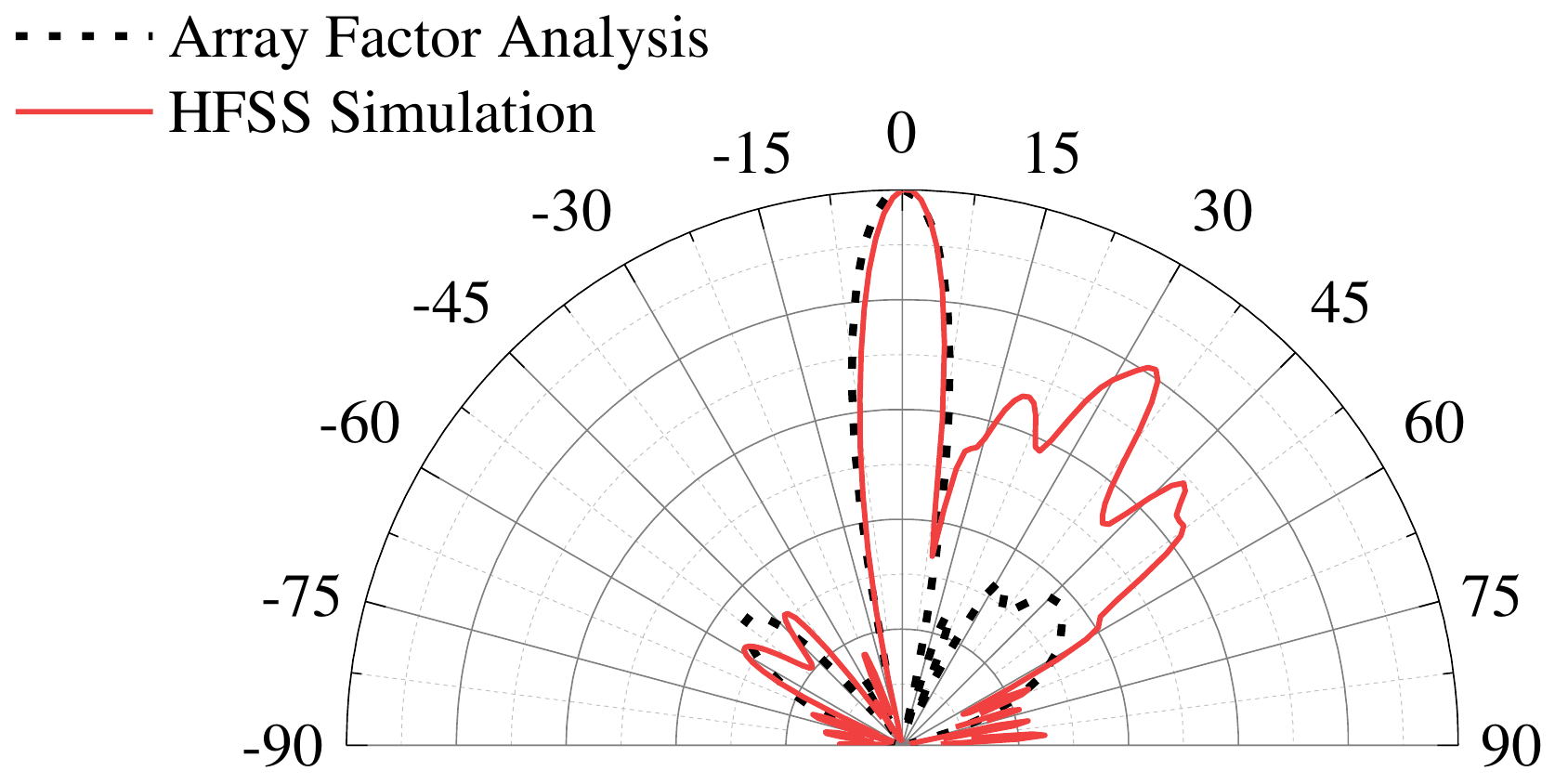}}
	\end{subfigure}
	\begin{subfigure}[At a Reflection Angle of $17^{\circ}$]
		{\includegraphics[width=0.27\textwidth]{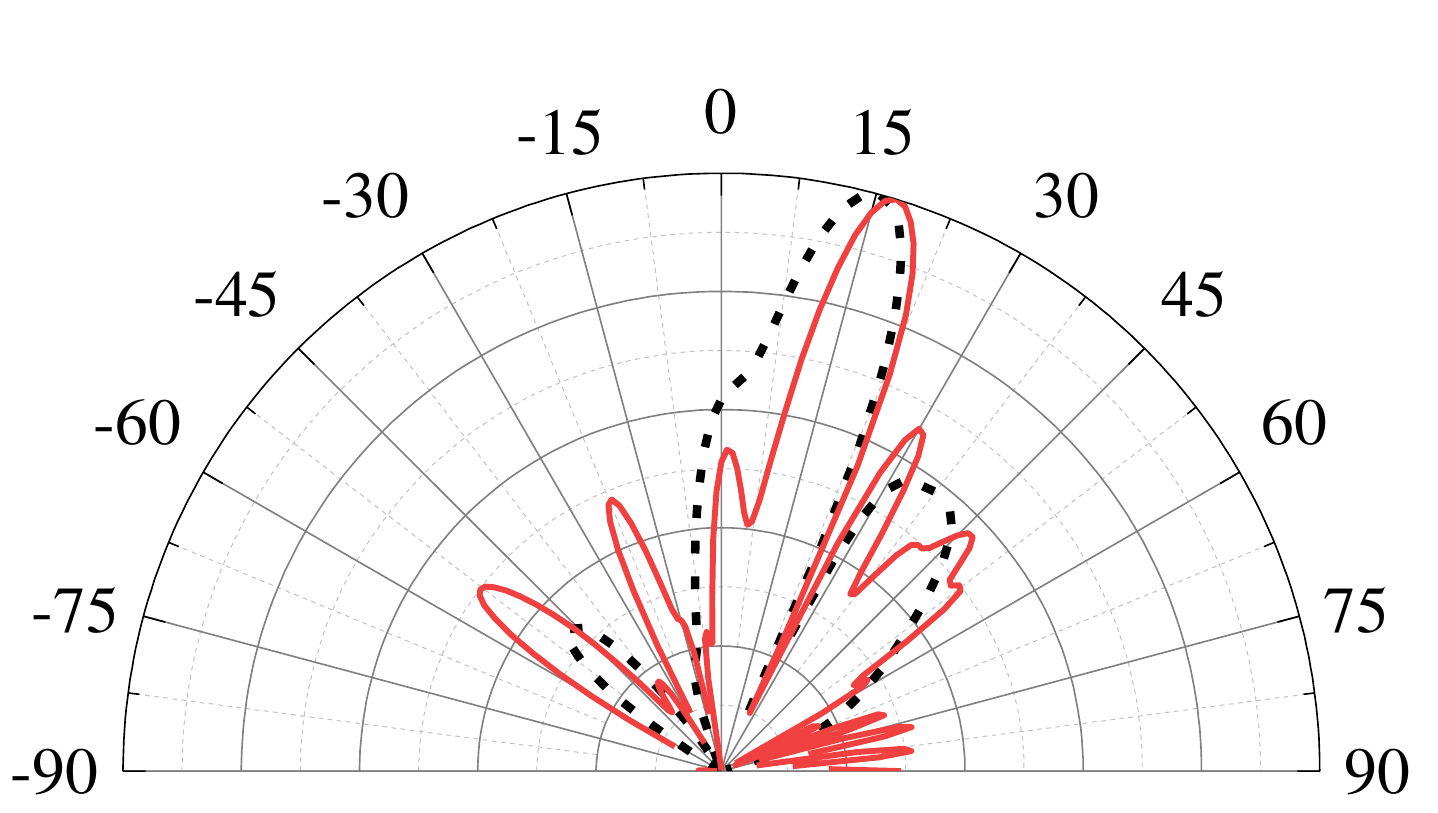}}
	\end{subfigure}
	\begin{subfigure}[At a Reflection Angle of $60^{\circ}$]
    	{\includegraphics[width=0.35\textwidth]{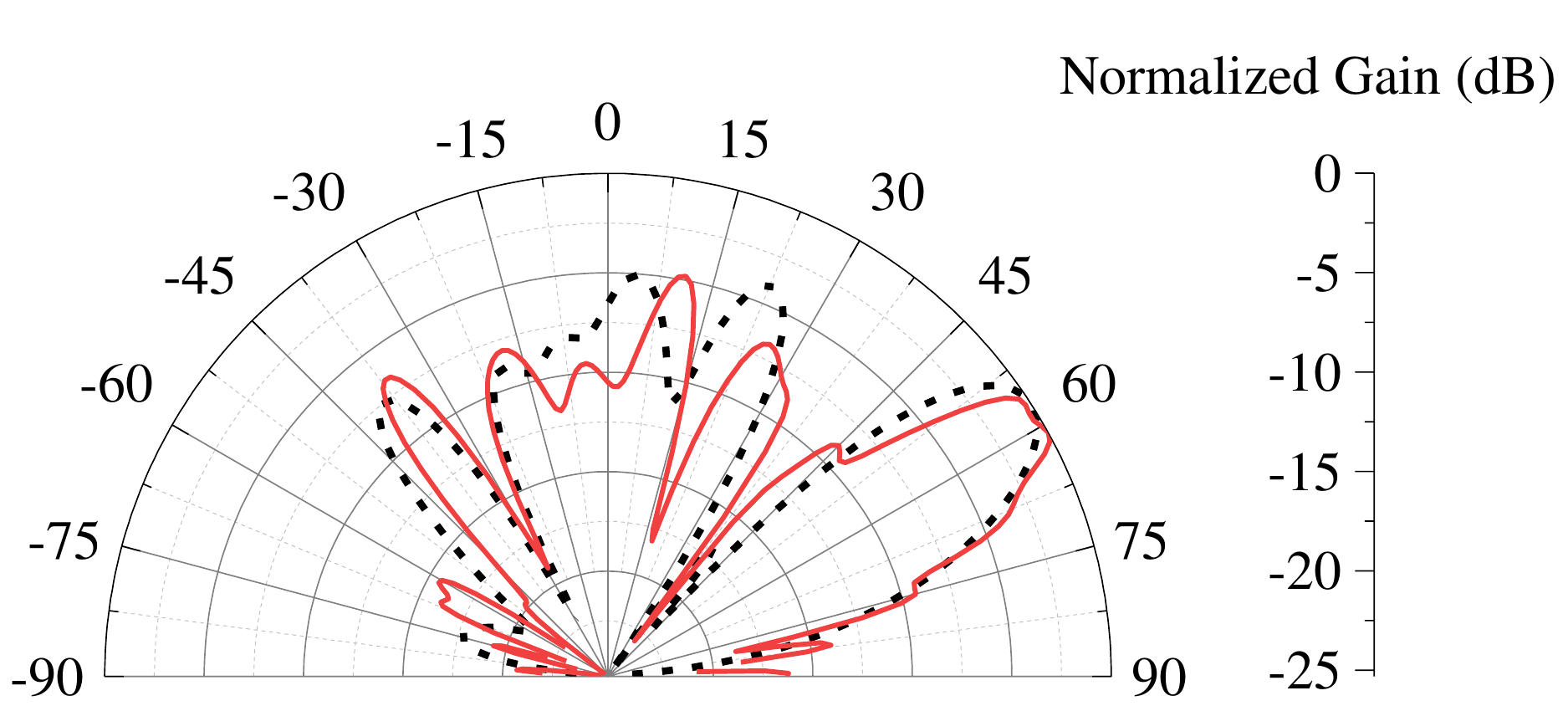}}
    \end{subfigure}
	\caption{Computed radiation patterns when the RIS is illuminated by a feed horn antenna (near-field, $\vartheta_{i}=-27.5^{\circ}$) for various reflection angles ($\vartheta_{d}$).} 
	\label{fig:sch_42}
\end{figure}

\begin{figure}[t]
	\centering
	\vspace{-3mm}
	\begin{subfigure}
		{\includegraphics[width=0.31\textwidth]{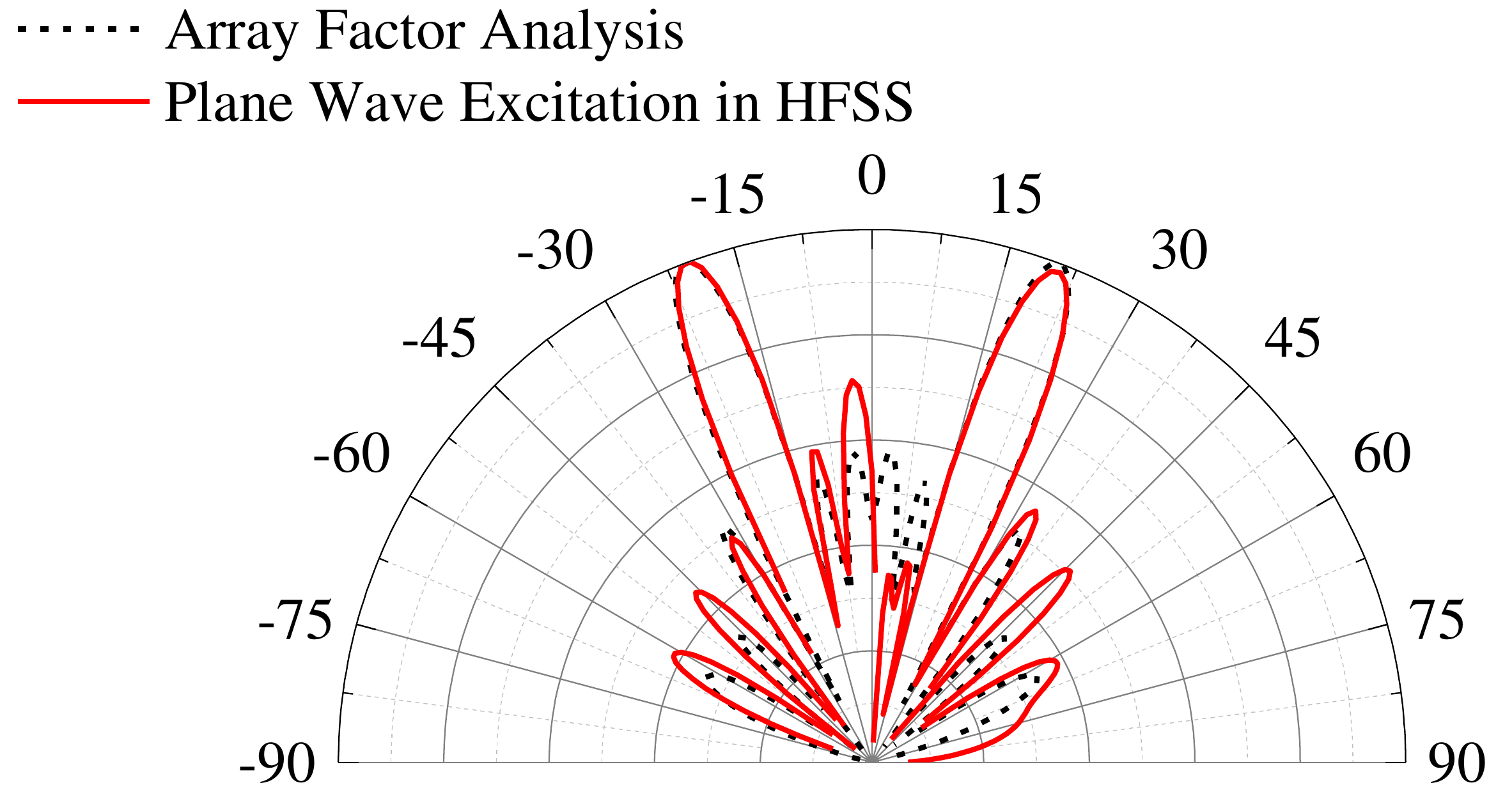}}
	\end{subfigure}
	\begin{subfigure}
		{\includegraphics[width=0.27\textwidth]{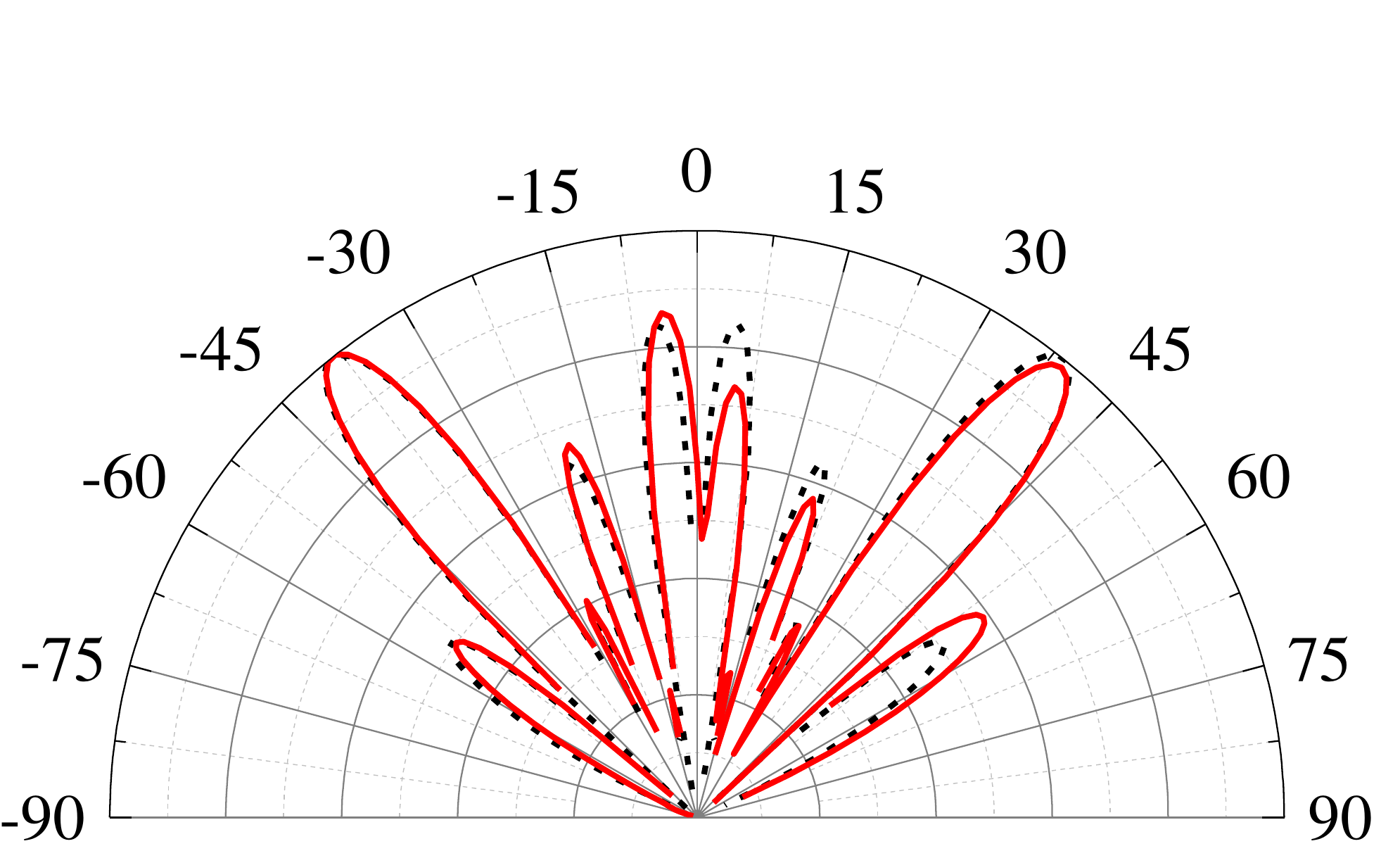}}
	\end{subfigure}
	\begin{subfigure}
    	{\includegraphics[width=0.35\textwidth]{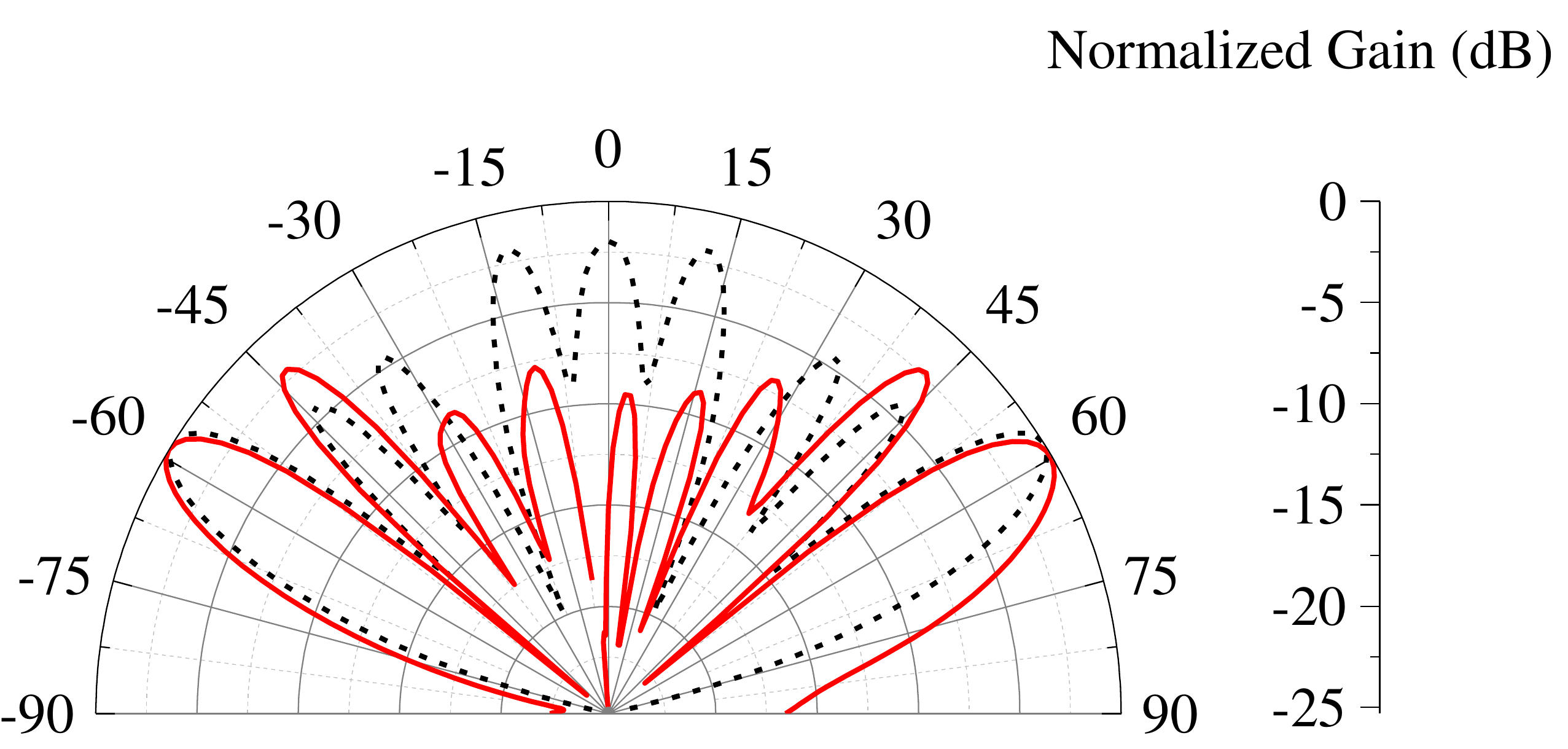}}
    \end{subfigure}
	\caption{Computed RIS radar cross section patterns under plane wave illumination (normal incidence) for various reflections angles ($\vartheta_{d}$): (a) $22.5^{\circ}$, (b) $40^{\circ}$, and (c) $60^{\circ}$.} 
	\label{fig:sch_42x}
\end{figure}

The radiation characteristics of the RIS have been studied using both the analytical expression of Eq. (14) as well as using a full-wave numerical method. To account for the electrically large model, we used ANSYS HFSS finite element boundary integral (FE-BI) which reduces the computational complexity. In the first scenario, we model the RIS with a feed horn antenna ($\theta_i = -27.5^{\circ}$) and calculate the radiation patterns for the excitation of three different reflection angles (0$^{\circ}$, 17$^{\circ}$, and 60 $^{\circ}$). The three excitations result in distinct main lobes at the desired directions, as shown in the analytical and full-wave simulation results of \figref{fig:sch_42}. The discrepancy between the patterns with respect to the sidelobe levels is attributed to the diffraction and surface wave phenomena that are not accounted for in the array factor analysis. Additionally, to  simulate  the beamforming characteristics when the feed is in the far-field, as expected in several wireless communications scenarios,  we replace the feed horn antenna by a plane wave excitation and plot the normalized radar cross section (RCS) pattern for three reflection angles (22.5$^{\circ}$, 40$^{\circ}$, and 60$^{\circ}$). Due to the phase quantization error, the reflected beam present a second grading lobe at the opposite angle around the specular direction, as shown in \figref{fig:sch_42x}. The quantization lobes can be mitigated by adding random phase delays at each unit cell \cite{kashyap_mitigating_2020, yin_single-beam_2020}.

\subsection{Characterization of a fixed beam RIS}

\begin{figure}[t]
    \centering
    \begin{subfigure}[]
        {\includegraphics[width=.35\columnwidth]{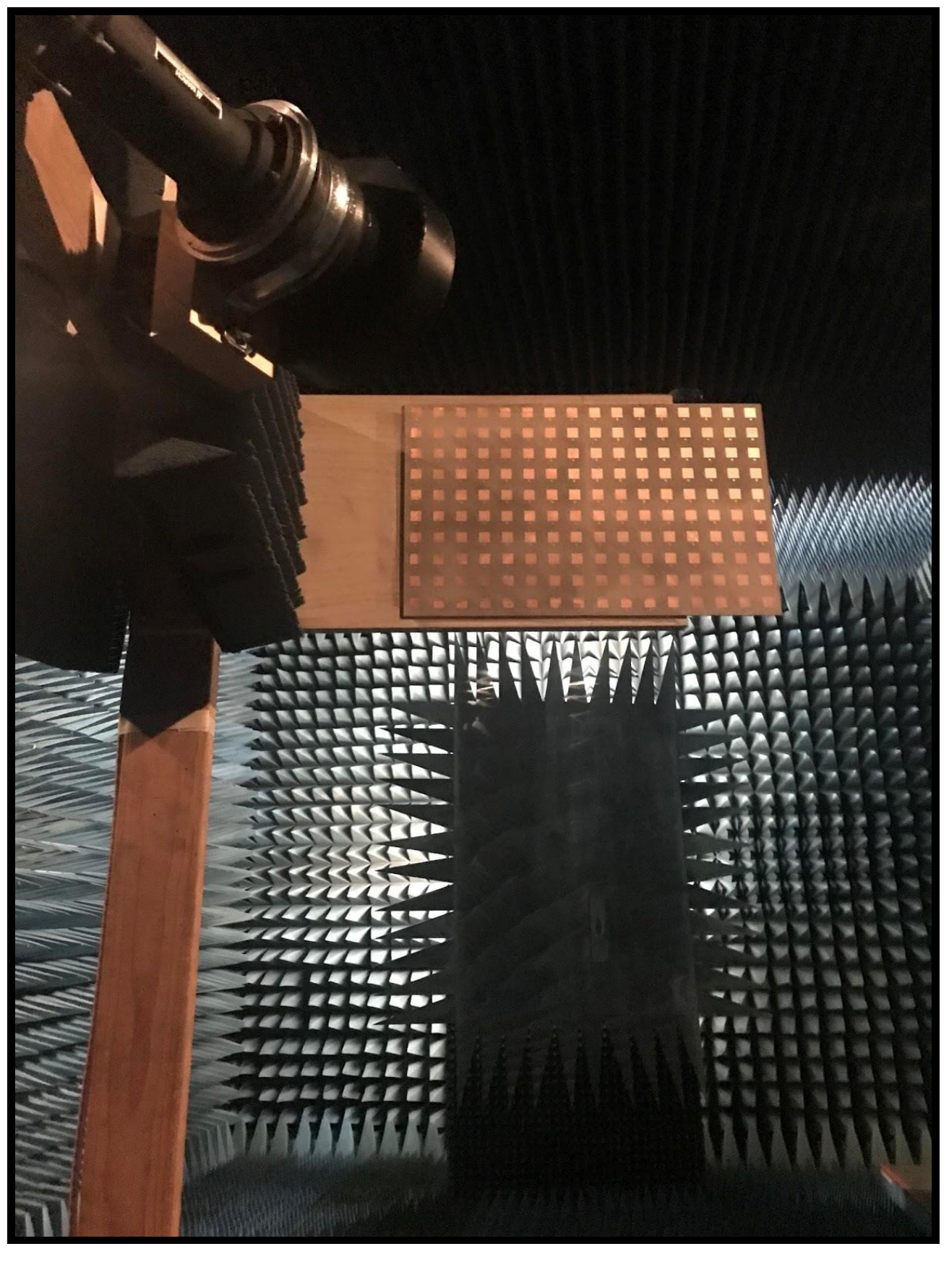}}
        \label{fig:Anechoic_Chamber}
    \end{subfigure} 
    \begin{subfigure}[]
     {\includegraphics[width = .46\columnwidth]{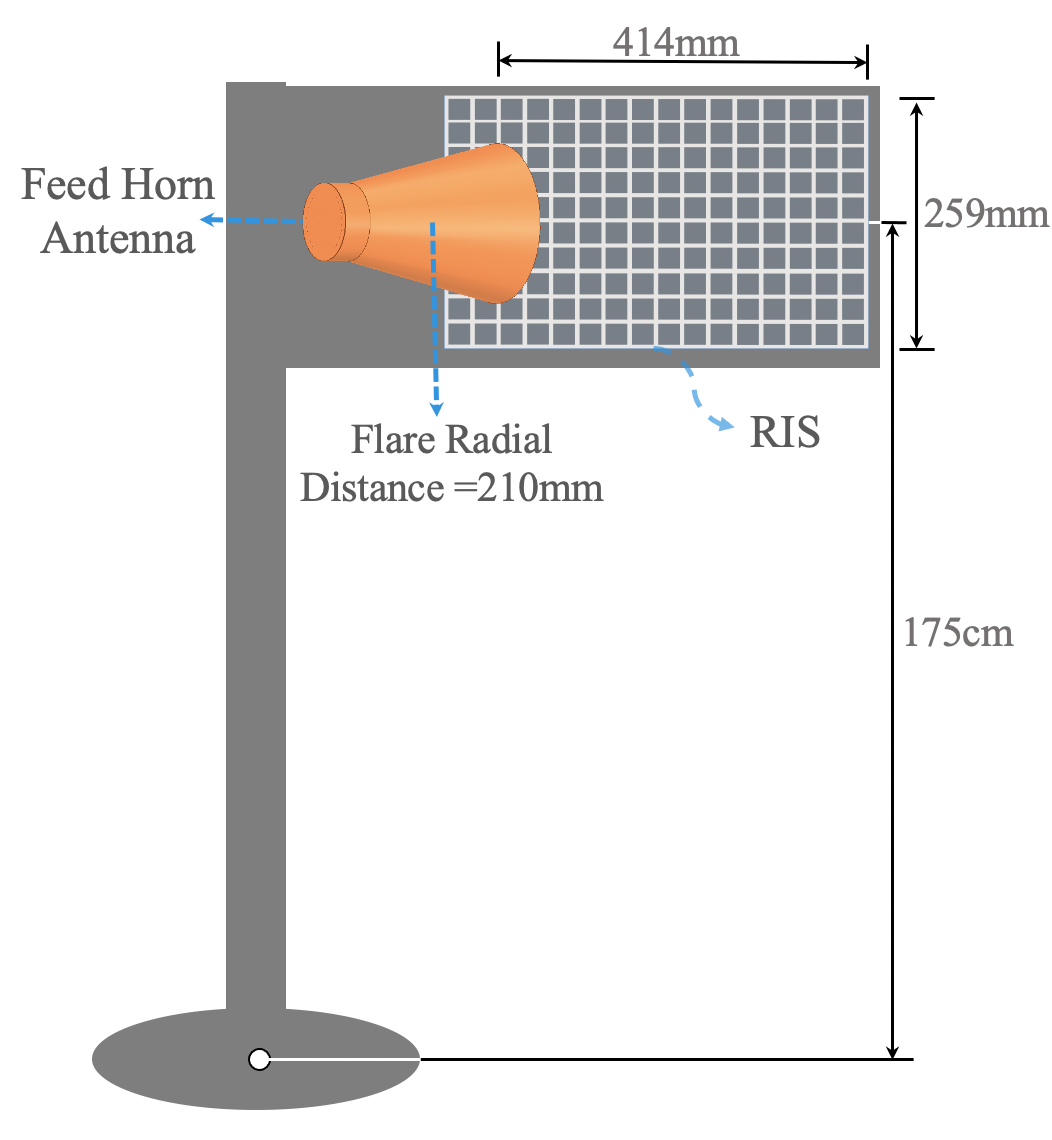}}
        \label{fig:Post_dimensions}
    \end{subfigure}
\caption{This figure illustrates the adopted measurement setup for the fixed-beam (passive) RIS prototype characterization. Figure (a) shows the anechoic chamber measurement setup and figure (b) summarizes the dimensions of the fixture.}
\label{fig:meas_chamber}
\end{figure}

\begin{figure}[t]
	\centering
	\includegraphics[trim= 0mm 160mm 0mm 30mm,clip,width=0.5\textwidth]{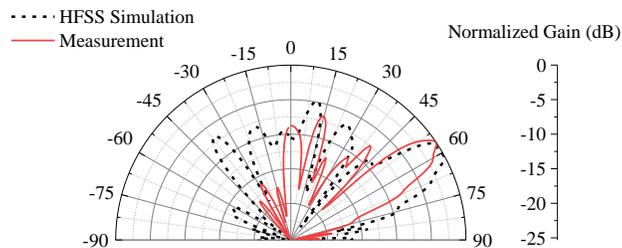}
	\caption{This figure shows a comparison between the measured (in the anechoic chamber) and computed radiation patterns of the 5.8 GHz fixed beam (passive) RIS prototype, which confirms the agreement between the designed and actual beams.}  
	\label{fig:sch_x45}
\end{figure}

To evaluate the accuracy of the array design simulations, we fabricated a fixed beam reflectarray and measured the radiation pattern at 5.8 GHz. The RIS comprises 160 unit cells (16 $\times$ 10) with an overall array dimension of 414 mm $\times$ 259 mm. In place of the RF PIN diodes, we place short or open terminations and design the RIS to operate as a reflectarray antenna with a reflection angle at $+60^{\circ}$. Then the layout was fabricated using a chemical etching process on a 2.54 mm thick Rogers RT/duroid 6002 substrate. First, the array layout is printed on a thermal sheet using an inkjet printer and then heat pressed on the substrate at 530 $C^{\circ}$. Afterward, the substrate with the imprinted mask is immersed into sodium persulphate ($Na_{2}S_{2}O_{8}$) to etch the copper and the mask is removed using acetone. To test the radiation pattern, a 12.5 dBi feed horn antenna is fixed above the reflectarray at an angle of -27.5$^{\circ}$, as depicted in \figref{fig:meas_chamber}. The measurement took place in an anechoic chamber (ASU's Compact Antenna Test Range) to ensure for minimum reflections from any surroundings. As plotted in  \figref{fig:sch_x45}, the normalized gain is measured in the [$-90^{\circ}$,$+90^{\circ}$] range ($E$-plane) and shows good agreement with the full-wave numerical analysis.

\begin{figure}[h]
	\centering
	\vspace{3mm}
	\begin{subfigure}
		{\includegraphics[scale=.5]{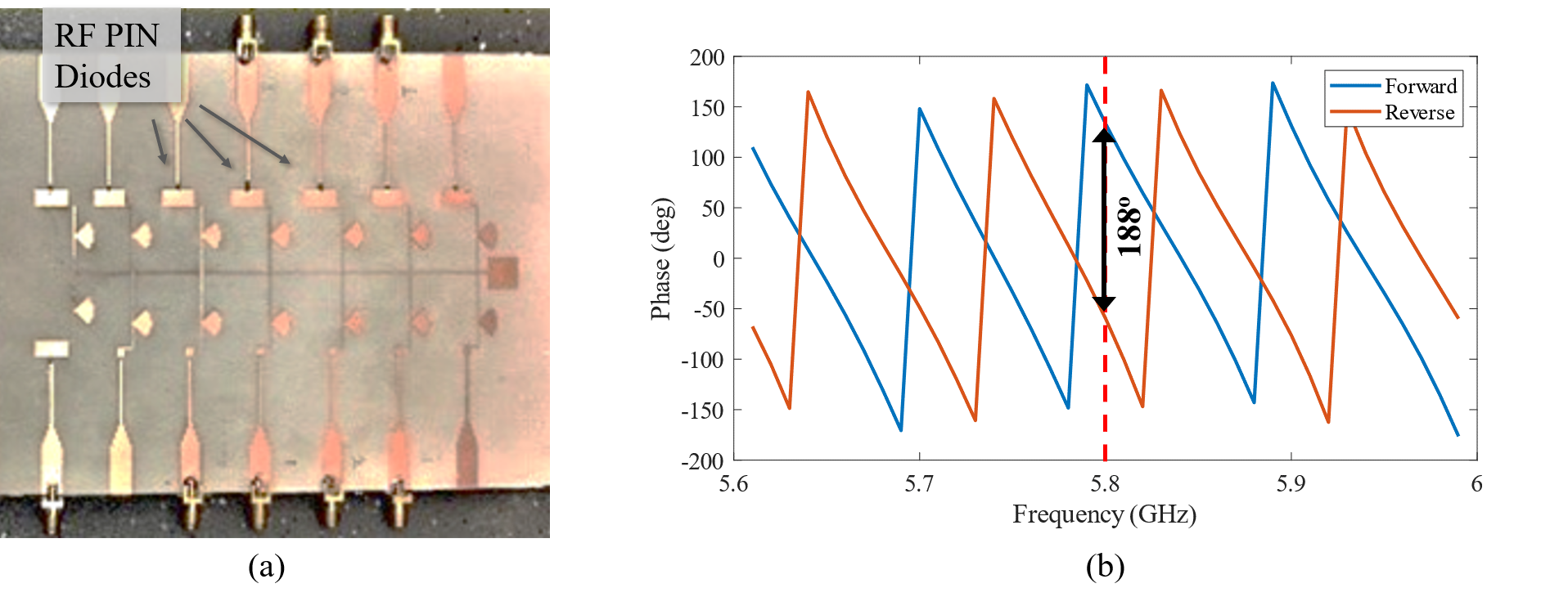}}
		\vspace{-6mm}
	\end{subfigure}
	
	\caption{Testboard for the characterization of the RF PIN diode switch. (a) Layout of the testboard and (b) phase response of the reflected signal for two different switch states (forward/reverse biased).}  
	\label{fig:sch_45}
\end{figure}

\begin{figure}
	\centering
	\includegraphics[scale = 0.4]{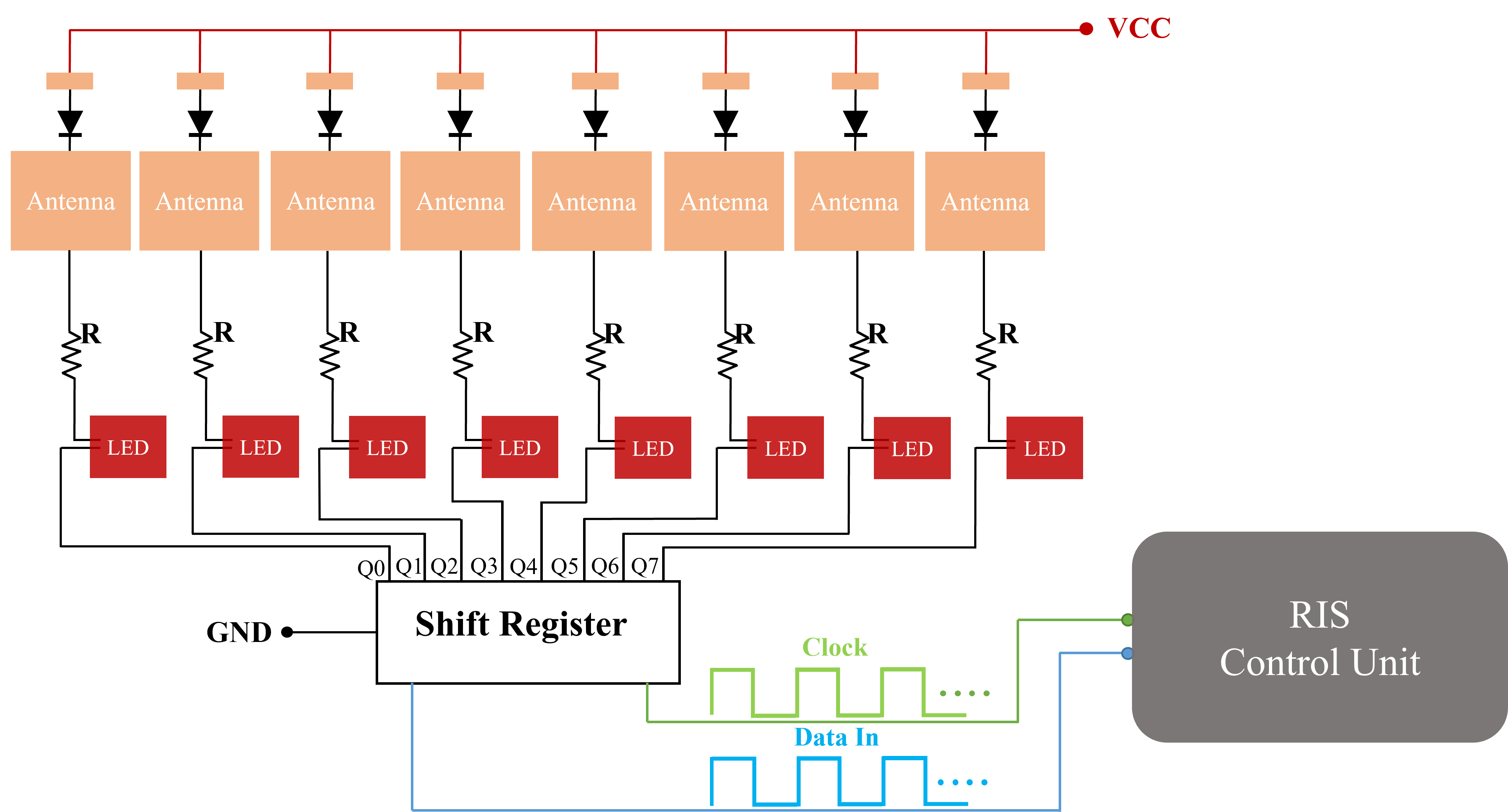}
	\caption{Schematic of the RIS control circuit. A codeword is generated and parsed to the RF PIN diodes through a microcontroller and 20 8-bit shift registers (LEDs are only needed for troubleshooting).}
	\label{fig:Biasing Circuit}
\end{figure}
\subsection {Implementation of the Reconfigurable Intelligence Surface}

\begin{figure}
\centering

    \begin{subfigure}[RIS Front-End]
        {\includegraphics[width = 0.5\linewidth]{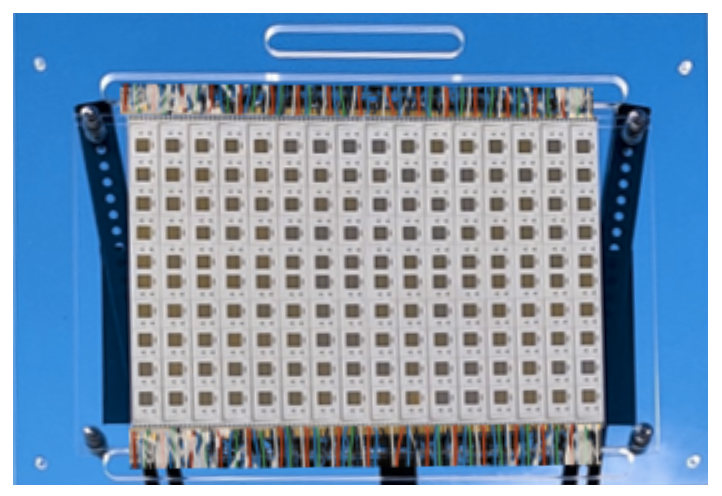}}
        \label{fig:RIS front}
    \end{subfigure} 
    \begin{subfigure}[RIS Back-End]
     {\includegraphics[width = 0.5\linewidth]{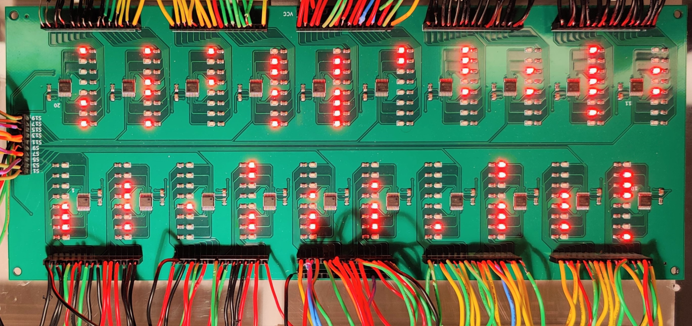}}
        \label{fig:RIS back}
    \end{subfigure}
\caption{The developed 160-element RIS prototype at the sub-6GHz band: Figure (a) shows the front-end which has the reflecting elements and figure (b) shows  the back-end with the RIS control circuit.}
\label{fig:RIS_setupf}
\end{figure}

In this section, we detail the fabrication and assembly of the RIS prototype. To verify the computed performance of the RF PIN diodes, we fabricated a test board that included several BAR50 PIN diodes with the necessary biasing contact pads and RF chokes, as shown in \figref{fig:sch_45}(a). Here, the diodes are terminated to a rectangular patch that is similar to the RIS unit cell topology, thus emulating a mismatch that generates around 180$^{\circ}$ phase difference in the reflected signal between the PIN diode's biasing states. Using a vector network analyzer (VNA), we measured the reflection coefficient ($S_{11}$) of 4 test diodes for both biasing states. As plotted in \figref{fig:sch_45}(b), the phase difference is around  $188^{\circ}$ for a wide frequency range around 5.8 GHz which was in accordance to the circuit simulations. Then, we assembled 160 diodes on the array and integrated them with the necessary control circuitry. \figref{fig:Biasing Circuit} shows a schematic of the RIS control circuitry that comprises the biasing circuit and a micro-controller (which also stores the  beam codebook). The micro-controller (Arduino MEGA2560) splits the bit sequence of each codeword (which consists of 160 bits) into 10 segments, each corresponding to the configuration weights of 16 unit cells. Every codeword segment is then pushed to a pair of in-series connected 8-bit shift registers (20 in total). As such, every output of a shift register is connected in series with an LED and resistor to provide 0.8 V across each PIN diode for forward biasing (0V for reverse biasing) and limit the current to 3 mA. We note here that the LEDs are used only for debugging and testing purposes and are not necessary for the RIS operation. With this biasing circuit design,  the bit sequence of each codeword determines the unit cell induced phased delay (0$^{\circ}$ or 180 $^{\circ}$) across the 160-element array, and thus steering the beam to the corresponding direction. \figref{fig:RIS_setupf} shows the front (antenna array) and rear side (control unit) of the RIS assembled on the fixture. The total DC power consumption of the RIS including the biasing circuitry is less than 0.4 Watts. Nevertheless, power consumption can be dramatically reduced with transistor-based switches  \cite{venkatesh_high-speed_2020} to micro-Watt levels even for RISs with thousands of elements due to the small leakage current of such devices. 

\subsection {Characterization of the RIS beamforming}
\begin{figure}[t]
 		\centering
 		\begin{subfigure}[]
 		{\includegraphics[width=0.35\linewidth]{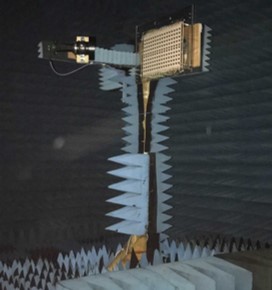}}
 		\end{subfigure}
 		\begin{subfigure}[]
 		{\includegraphics[width=.35\linewidth]{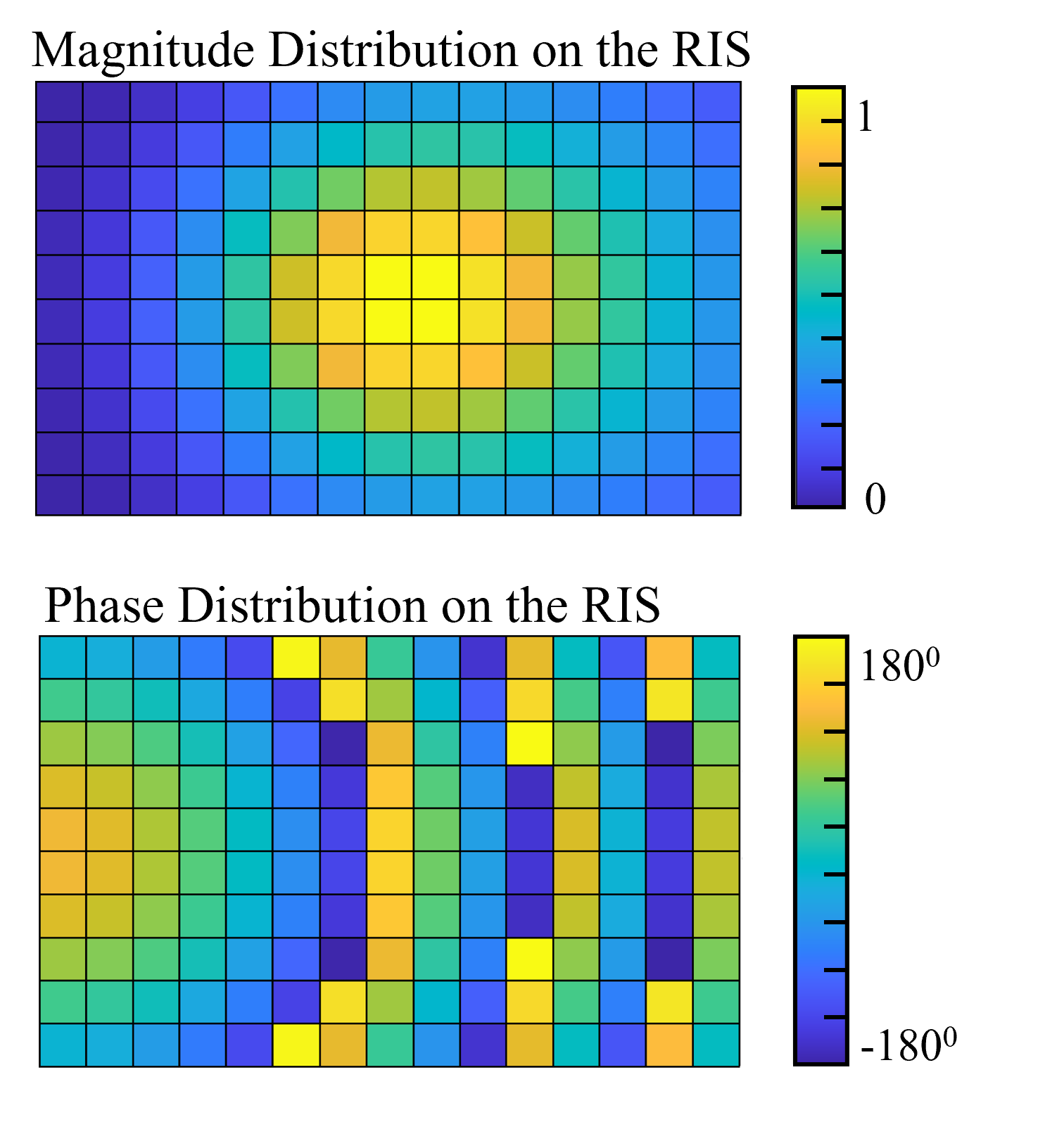}}
 		\end{subfigure}
 		\caption{RIS characterization in the anechoic chamber. (a) Photo of the RIS setup in the chamber and (b) magnitude and phase distribution of the incident feed horn antenna beam on the RIS aperture.} 
 		 \label{fig:chamber setup}
 	\end{figure}

\begin{figure}
\centering
    \includegraphics[scale = 0.6]{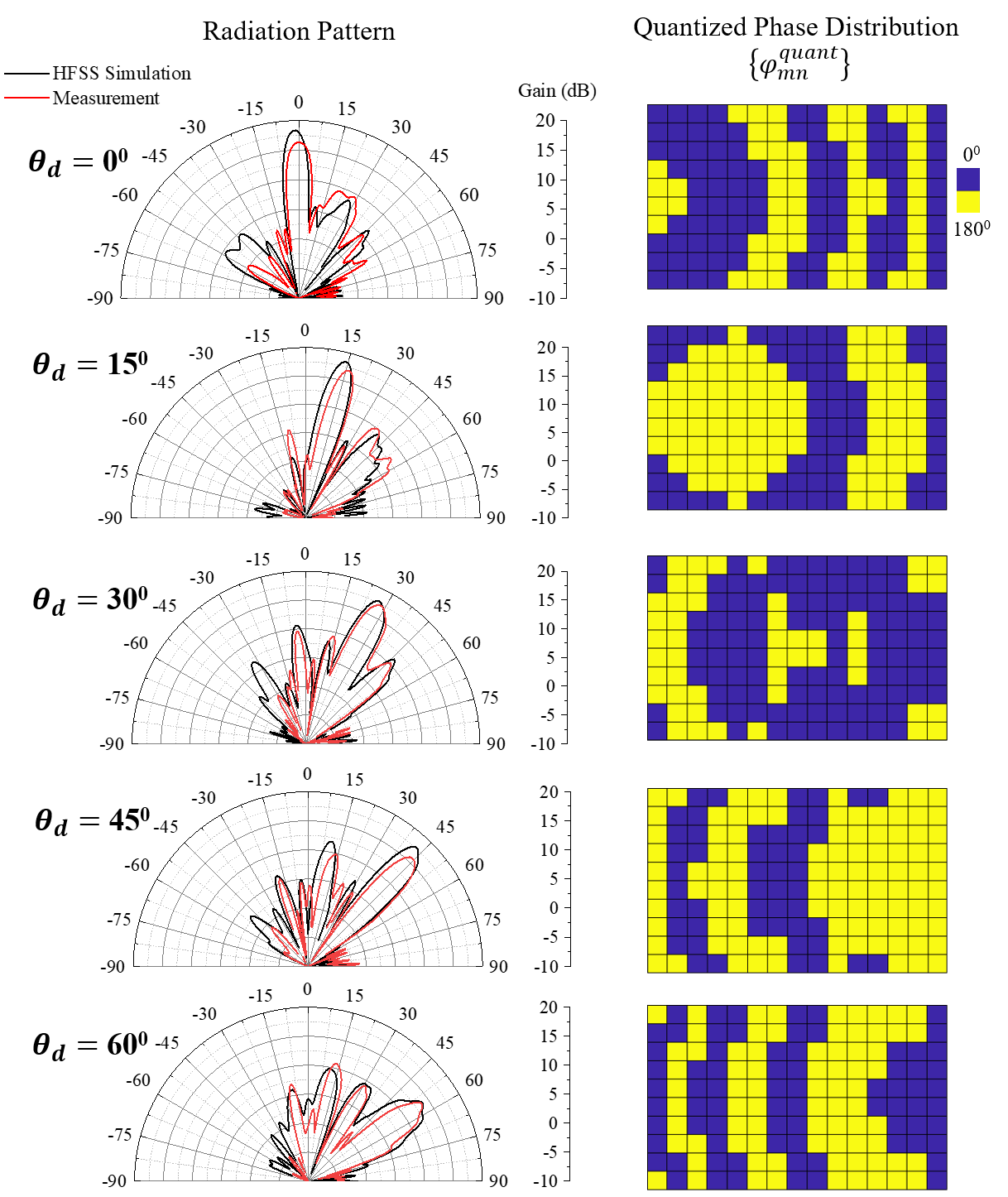}
    \caption{Characterization of the RIS beamforming for various reflection angles $\theta_{d}$ at 5.8 GHz. Left column: Comparison between computed and measured radiation patterns. Right column: Quantized (1-bit) phase shift distribution $\Phi^{quant}$ on the RIS surface to generate the respective radiation patterns.}
    \label{fig:Rad_Pat_Comp}
\end{figure}

\begin{figure}
\centering
    \includegraphics[scale = 0.35]{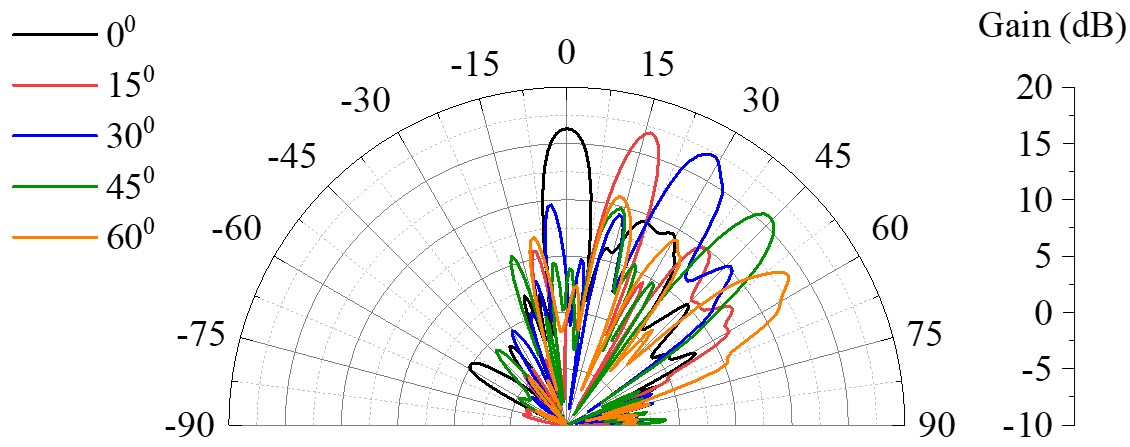}
    \caption{Comparison of RIS measured radiation patterns for five $\theta_{d} $ angles.}
    \label{fig:Meas_Rad_Pat}
\end{figure}

Similar to the fixed beam reflectarray measurement, the RIS is assembled with a feed horn antenna (near-field) inside the anechoic chamber, as shown in \figref{fig:chamber setup}(a). Being in the near-field, the feed horn impresses a Gaussian field distribution on the RIS aperture, as shown in the plots of magnitude and phase in \figref{fig:chamber setup}(b). The recorded  radiation patterns for 5 reflected directions ($\vartheta_{d}$ = $\{$ 0$^{\circ}$, 15$^{\circ}$, 30$^{\circ}$, 45$^{\circ}$, 60$^{\circ}$ $\}$) show very good agreement with the full-wave simulations, as plotted in  \figref{fig:Rad_Pat_Comp}. The right column in \figref{fig:Rad_Pat_Comp} shows the quantized phase distribution $\Phi^{quant}= ( \phi_{mn}^{quant} ) $ when applying the respective codeword vectors $\overline{\boldsymbol{\psi}}_{-27.5^{\circ},d}$ on the RIS switches. Finally, \figref{fig:Meas_Rad_Pat} compares the measured scanned angles on the same plot. The measurements confirm the capability of the RIS to scan in the $ \pm 60^o $ range maintaining a single main lobe. Additionally, the side lobe level (SLL) is maintained below -7 dB for more than 100 MHz bandwidth (5.75-5.85 GHz) and below -5 dB for more than 400 MHz at $\theta_d= 30^{\circ}$.

\section{Integrating the RIS into a Wireless Communication Testbed}

\begin{figure}
	\centering
	\includegraphics[width=.85\columnwidth]{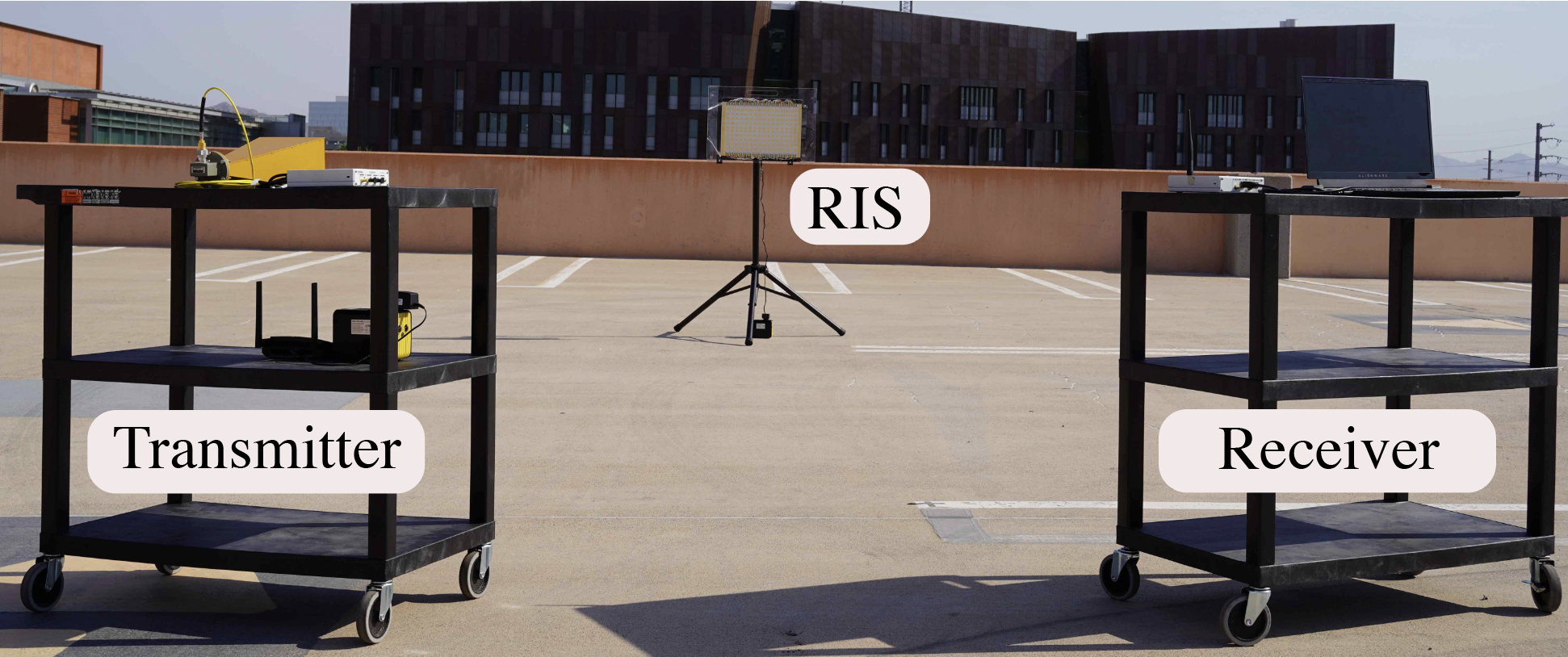}
	\caption{The RIS-assisted wireless communication system consists of an RIS, a  transmitter (BS), and a  receiver (UE). The transmitter and receiver are portable and can be placed in the far-field of the RIS surface.}
	\label{fig:setup}
\end{figure}

To demonstrate the potential of reconfigurable intelligent surfaces, we integrate the fabricated prototype, described in \sref{sec:design}, into a wireless communication testbed. In this section, we describe the developed testbed and beam selection process.

\subsection{Testbed Description}  \label{sec:testbed}
As illustrated in \figref{fig:setup}, our RIS-integrated wireless communication testbed consists of (i) a single-antenna transmitter, (ii) a single-antenna receiver, and (iii) the RIS. Next, we summarize the key aspects of the testbed
\begin{itemize}
	\item \textbf{Transmitter:}  We consider a single-antenna transmitter that is implemented using an NI USRPs 2901 operating at 5.8GHz. The USRP is connected to a  horn antenna with a 18.5 dBi gain, emulating the antenna gain of a BS. As will be explained shortly, the transmitter operation is controlled by a Raspberry Pi that is wirelessly controlled (over a 2.4GHz channel). 
	
	\item \textbf{Receiver:} Similar to the transmitter, the communication receiver is implemented using an NI USRP 2901 operating at the 5.8GHz band. The USRP is connected to either a dipole antenna or a horn antenna with 12.5 dBi gain. The receiver is controlled by a laptop that manages the overall operation of the RIS-integrated wireless communication system.
	
	\item \textbf{RIS:} The developed RIS, consisting of 160 elements (16 $\times$ 10) is placed and leveraged in reflecting the transmitted signal to the receiver direction.  Section \ref{sec:design} provides a detailed description for the design and fabrication of the adopted RIS.  
	
\end{itemize} 

The developed RIS-integrated wireless communication system operates at the sub-6GHz band with a center frequency of $5.8$GHz and a $20$MHz bandwidth. The system adopts an OFDM transmission/reception with $64$ subcarriers. In the next subsection, we describe the operation framework including the selection of the reflection beamforming codewords at the RIS surface.  

\subsection{Testbed Operation and Beam Selection}

The main objective of the developed testbed is to evaluate the coverage gains when using an RIS. Toward this objective, we adopt the following operation framework. For given locations of the transmitter and receiver, the central controller triggers the transmitter to send OFDM-based pilot sequences. During this transmission, the controller wirelessly orders the Raspberry Pi  that is controlling the RIS configurations to switch between the $N_i N_d$ beams in the codebook $\boldsymbol{\mathcal{P}}$. With this beam training, the receiver measures the power of the received signal $r_k$, from equation \eqref{eq:rec}, and selects the optimal RIS configuration $\boldsymbol{\psi}^\star$ that solves 
\begin{equation}
		{\overline{\boldsymbol{\psi}}}^\star =  \arg \hspace{-1pt}  \max_{	\overline{\boldsymbol{\psi}} \in \boldsymbol{\mathcal{P}}} \ \ \ \ \sum_{k=1}^K {\left| r_k\left(\overline{\boldsymbol{\psi}}\right) \right|^2},
\end{equation}
where $r_k\left(\overline{\boldsymbol{\psi}}\right)$ is the received signal at the k-th subcarrier when the reflection beam codeword $\overline{\boldsymbol{\psi}}$ is used by the RIS. For the  1-bit codebook, $\boldsymbol{\mathcal{P}}$, we follow the approach described in \sref{subsec:codebook} to design it. The exact angular range and number of beams for the codebook adopted in the field measurements are provided in \sref{sec:Results}.

\section{Field Tests and  Results} \label{sec:Results}
In this section, we present the results of our field tests that evaluate the beamforming capabilities and coverage improvement of the developed RIS-based wireless communication system. First, we describe the setups adopted in the measurement campaigns in Subsection \ref{subsec:setup} before demonstrating the results in the following subsections. \textbf{It is worth mentioning  that all our measurement campaigns are conducted outdoors to account for wave propagation phenomena that  occur in real-world scenarios, including scattering from the terrain, posts, and edge diffraction. Therefore, the results of these campaigns draw important insights into the actual performance of RISs in  realistic wireless communication deployments. }

\subsection{Measurement  Setups} \label{subsec:setup}
In this subsection, we describe the two adopted setups, namely the ASU parking lot and the ASU Gammage Memorial Auditorium, where our measurements are performed. 
\begin{figure}[t!]
\centering
    \includegraphics[scale =.55]{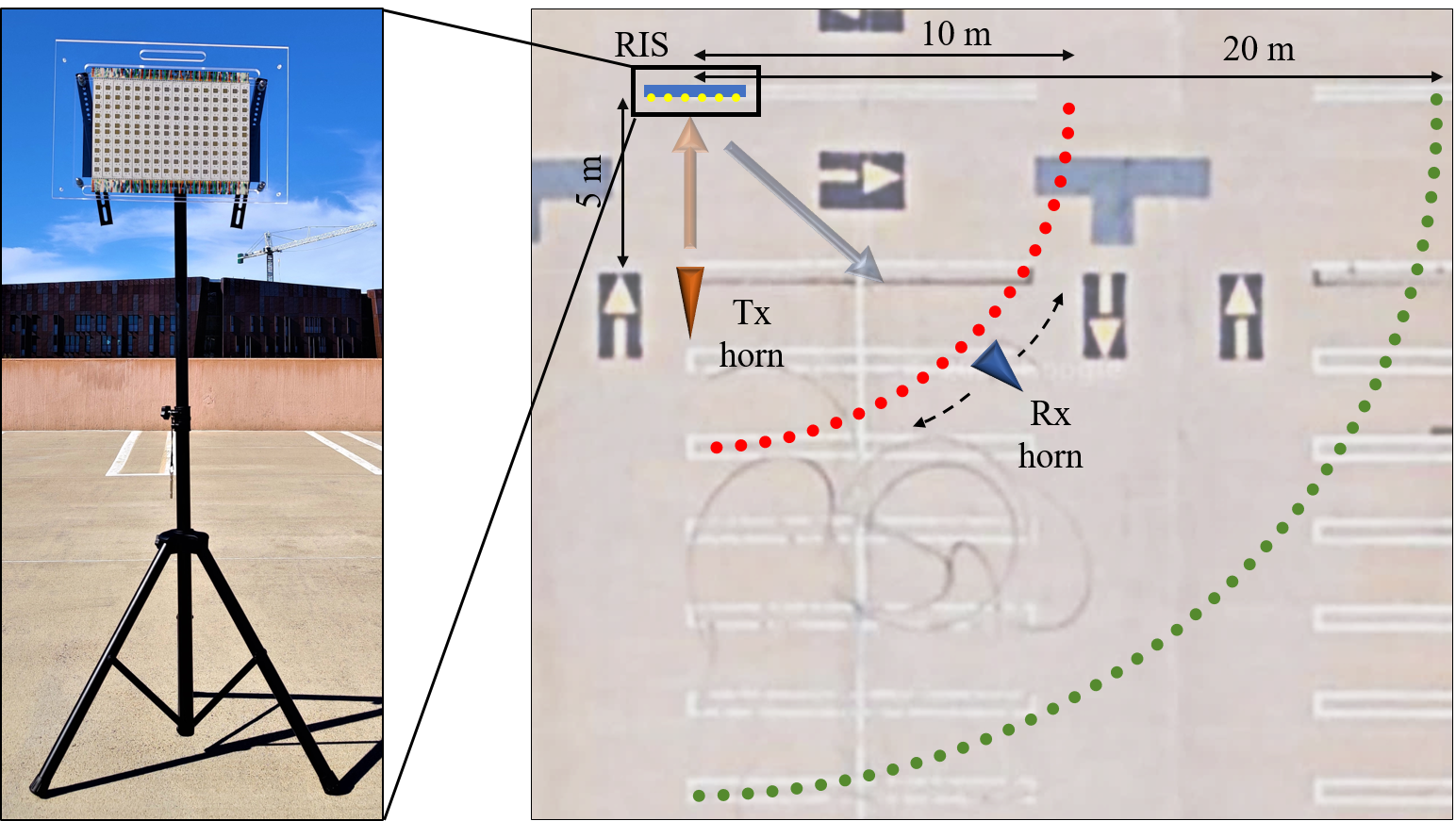}
    \caption{Measurement Setup 1 (ASU Campus Parking Lot). To test the RIS' electronic beamscanning capabilities in the field, the RIS is illuminated from the boresight direction ($\vartheta_{i}=0$) and the reflected signal is recorded at various angles and distances. [Inset: RIS mounted on a the tripod.]}
    \label{fig:Beamscanning Meas}
\end{figure}

\textbf{Measurement Setup 1: ASU Campus Parking Lot:}
To characterize the beamforming codebook in an LoS real-world scenario, we carried out the measurements in a parking lot. The satellite image of the parking lot with the measurement setup overlaid is shown in Fig.~\ref{fig:Beamscanning Meas}. The setup consists of the prototype-RIS mounted on a tripod, a C-band corrugated conical horn antenna with a gain of 12.5 dBi used as the feed, and a C-band pyramidal horn with a gain of 18.5 dBi used on the receiver. The feed and the receiver horns were respectively connected to transmitting and receiving USRP modules. The feed horn is positioned in front of the RIS at 5m distance from the surface. The tripod-mounted-RIS is depicted in the inset of Fig.~\ref{fig:Beamscanning Meas}. The receiver is positioned at a distance of 10 m in front of the the RIS and rotated in a circular arc from 0$^{\circ}$-60$^{\circ}$ to capture the deflected signals from the RIS. Further, to characterize the pathloss, the measurements were repeated at 20 m and 40 m from the RIS.

\begin{figure}
    \centering
    \includegraphics[width = .65\columnwidth]{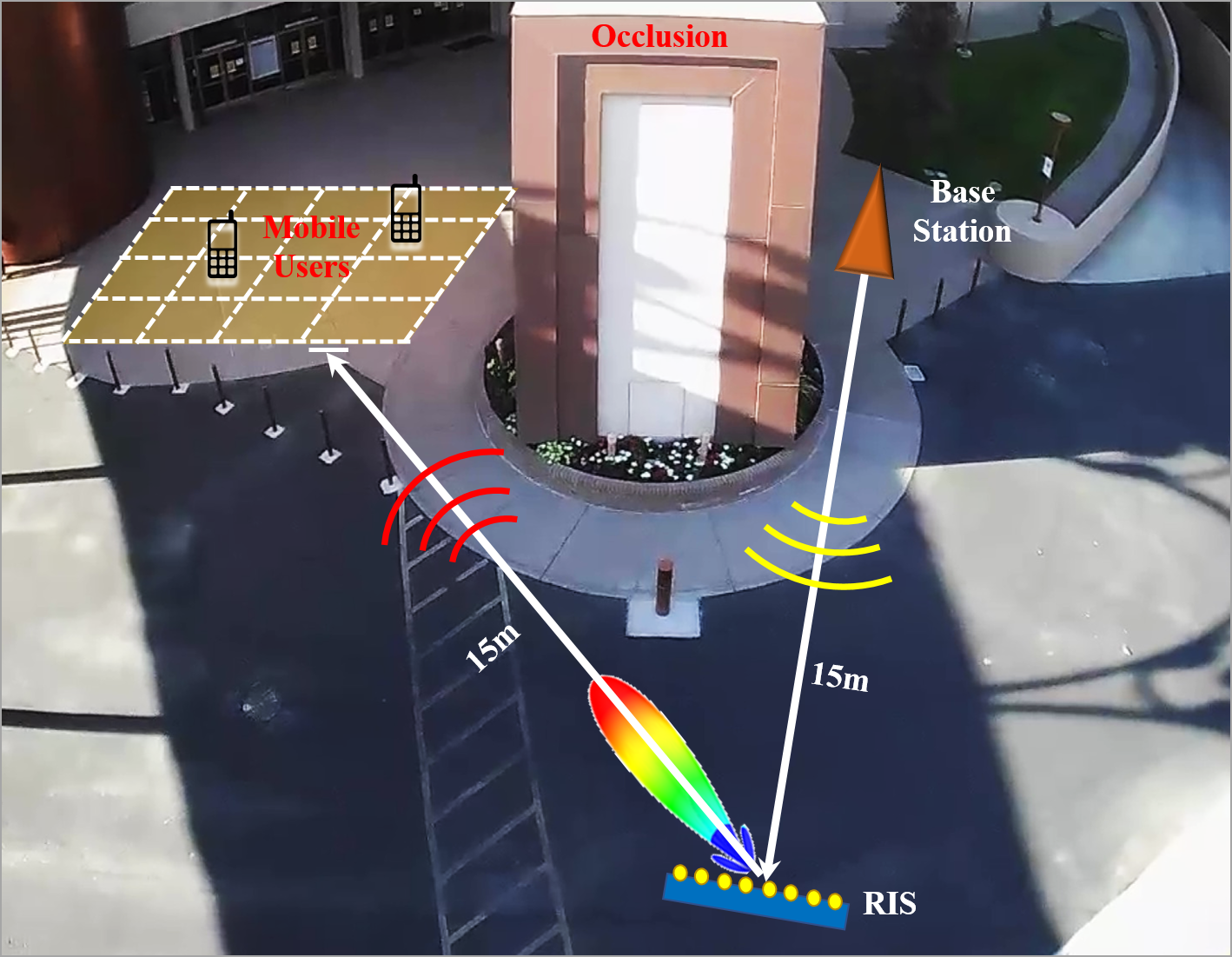}
    \caption{Measurement Setup 2 (ASU Gammage Memorial Auditorium). In this field test, a strong occlusion (blockage) exists between the BS and the mobile user. The RIS is strategically placed to leverage its  beamscanning capabilities in extending the coverage to the LoS-obstructed mobile users.}
    \label{fig:Signal Coverage}
\end{figure}

\textbf{Measurement Setup 2: ASU Gammage Memorial Auditorium:}
To emulate a user environment with significant signal degradation, we carried out the coverage measurements around the entrance of ASU's Gammage Memorial Auditorium, shown in Fig.~\ref{fig:Signal Coverage}. Here, the base station is modeled using a 19dBi horn antenna placed on one side of a tall (5m) and thick (2m) concrete wall, covering the north and eastern side of the outdoor area of the venue. On the other side of the wall, the receiver uses an ommi-directional antenna and moves in the area in front of the venue entrance. The concrete wall serves as an occlusion between the transmitter and the receiver, thus we examine the level of the received signal with and without the use of the RIS.

\subsection{Characterization of the Beamforming Codebook}

\begin{figure}[t!]
	\centering
	{ \includegraphics[width=.7\columnwidth]{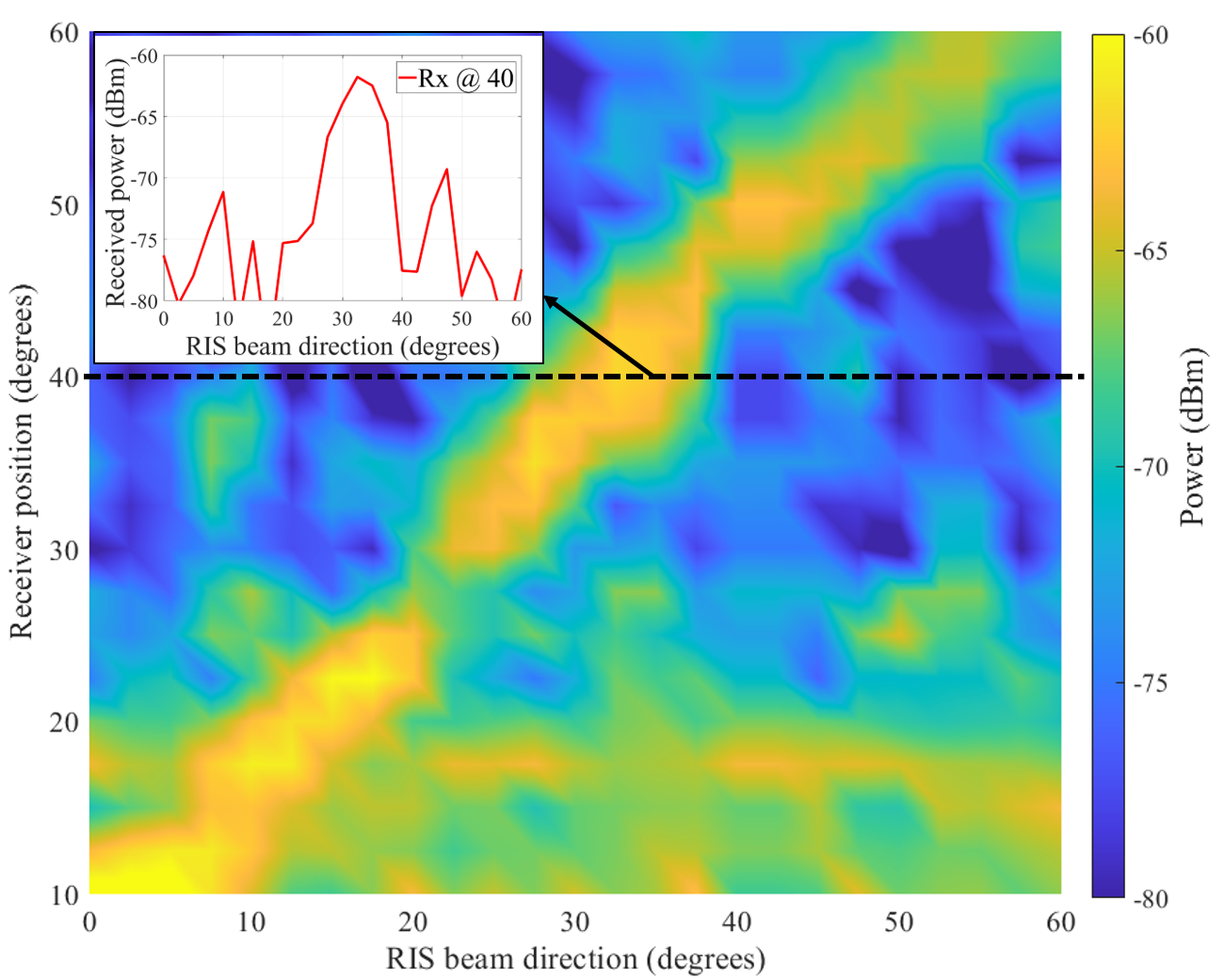}}
	\caption{Evaluating the RIS' electronic beamscanning (azimuth plane). The high receive power at the diagonal line confirms the agreement between the codebook beam directions and the receiver location (which is moving along a circular path, as shown in \figref{fig:Beamscanning Meas}). [Inset: 2D plot of the received signal when the receiver is oriented 40$^{\circ}$ from the broadside] }
	\label{fig:Beamforming data}
\end{figure}

Here, we present the results obtained from the field measurements carried out to characterize the beam-scanning capabilities of the  proposed RIS.
A 20 MHz bandwidth OFDM signal centered around 5.8 GHz is coupled to the feed horn from the transmitting USRP module. The feed horn is aligned in front of the RIS such that it illuminates the RIS from broadside. The reflected signals are recorded using a receiver horn antenna along a 10$^{\circ}$-60$^{\circ}$ arc with a radius of 10 m at every 2.5$^{\circ}$. After beam-scanning at every point, we form a reception pattern for every scanning RIS scanning angle. The 3D surface plot of Fig.~\ref{fig:Beamforming data} shows the beamforming capability of the RIS proposed in this work. As the beams from the RIS are scanned from 0$^{\circ}$-60$^{\circ}$ employing the beamforming codebook, the peak power is received at the corresponding receiver location. The received power at an exemplary angle of 40$^{\circ}$ from the broadside is shown in the inset of Fig.~\ref{fig:Beamforming data}. For receiver angles less than 20$^{\circ}$, coupling is noticed between the transmitter and the receiver. This is primarily due to the transmitter's backlobes as well as its close proximity to the receiver for directions near broadside. This is an artifact due to the limitations of the measurement setup and would not be present in a deployment where the transmitter is much farther or blocked by an occlusion. Nevertheless, the power corresponding to the the desired direction is still sufficiently high to offset this coupling effect and the receiver is able to detect the RIS beam. \textbf{Overall, Fig.~\ref{fig:Beamforming data} shows that the developed RIS can achieve around $18$-$20$dB SNR gain in the considered deployment scenario.} It is worth mentioning here that these beamforming gains can potentially be  improved if these codebooks are further optimized to match the RIS hardware impairments and the surrounding environment \cite{zhang2021reinforcement,alrabeiah2020neural}.

\subsection{Pathloss Measurements}

\begin{figure}
	\centering
	{ \includegraphics[width=.7\columnwidth]{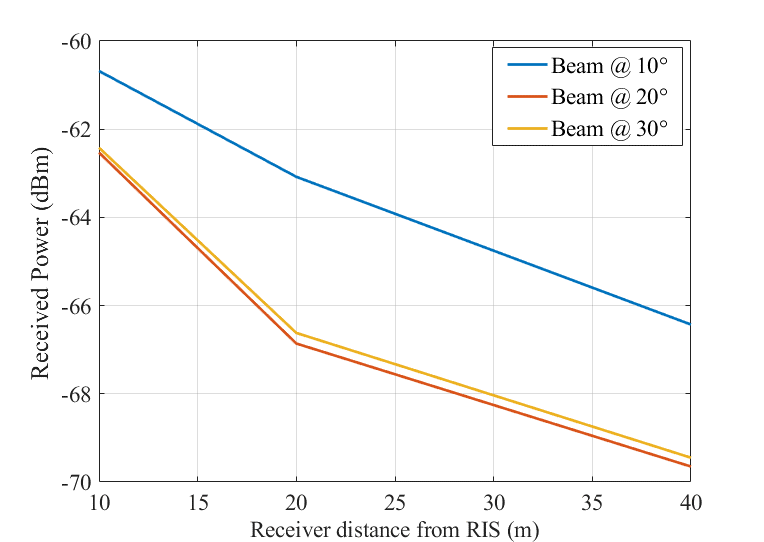}}
	\caption{Path loss as a function of distance: The receiver is fixed at various directions ( $\theta_d = {10^{\circ}, 20^{\circ}, 30^{\circ}} $) from broadside and the distance between the LIS and the receiver is increased from 10 m to 40 m.}
	\label{fig:Path loss}
\end{figure}

To characterize the pathloss, similar beamforming measurements were repeated at distances of 20 m and 40 m from the RIS. Fig.~\ref{fig:Path loss} depicts the change in the received power levels, at few nominal angles (10$^{\circ}$, 20$^{\circ}$ and 30$^{\circ}$) from broadside, as a function of distance from the RIS. As expected, the pathloss increases as the distance from the RIS increases as well as when the beam deviates away from the broadside direction.

\subsection{Signal Coverage Measurements}

\begin{figure}
    \centering
    \begin{subfigure}[Without RIS]
        {\includegraphics[width = 0.49\columnwidth]{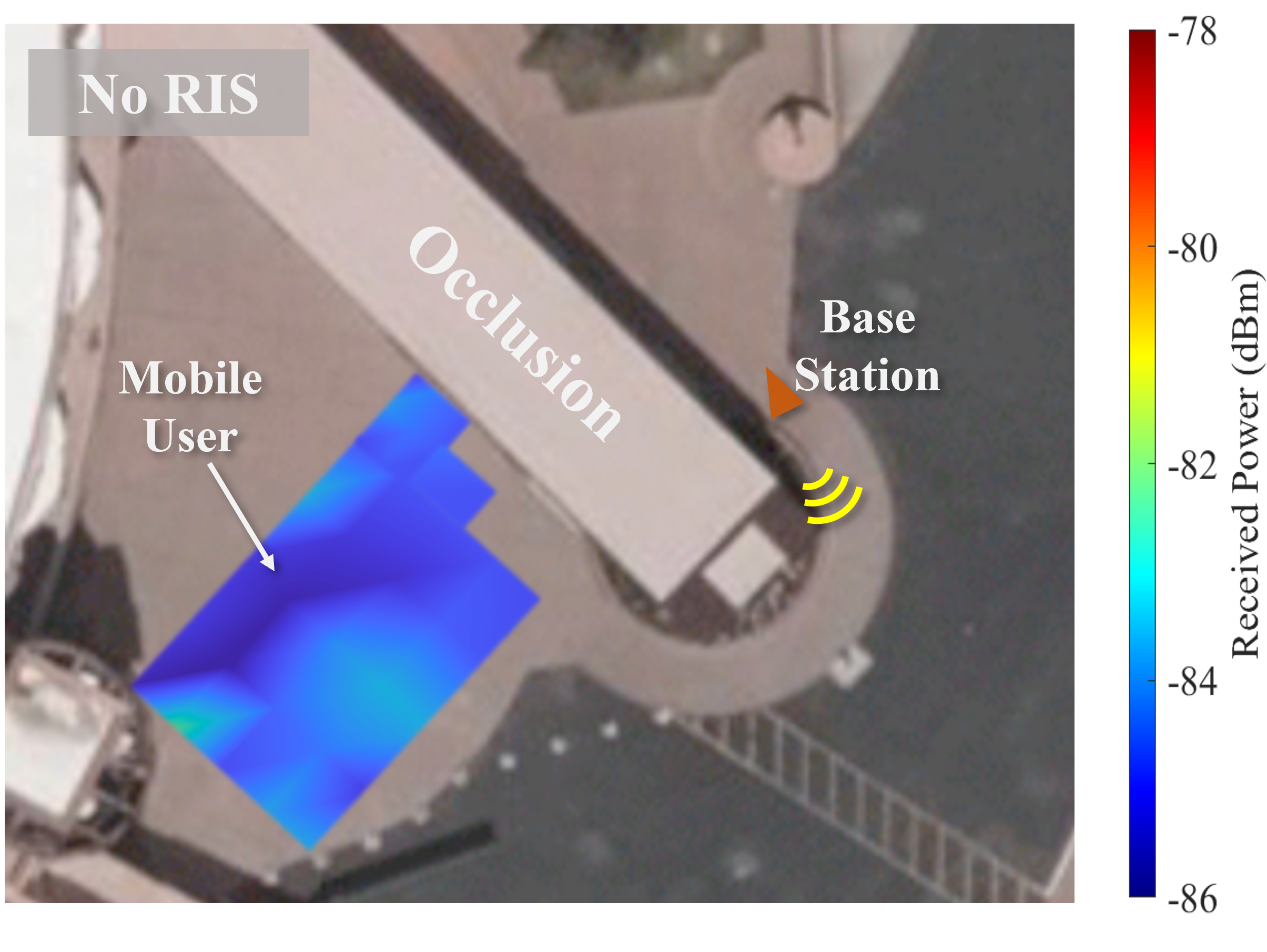}}
        \label{fig:no RIS}
    \end{subfigure} 
    \begin{subfigure}[With RIS]
        {\includegraphics[width = 0.49\columnwidth]{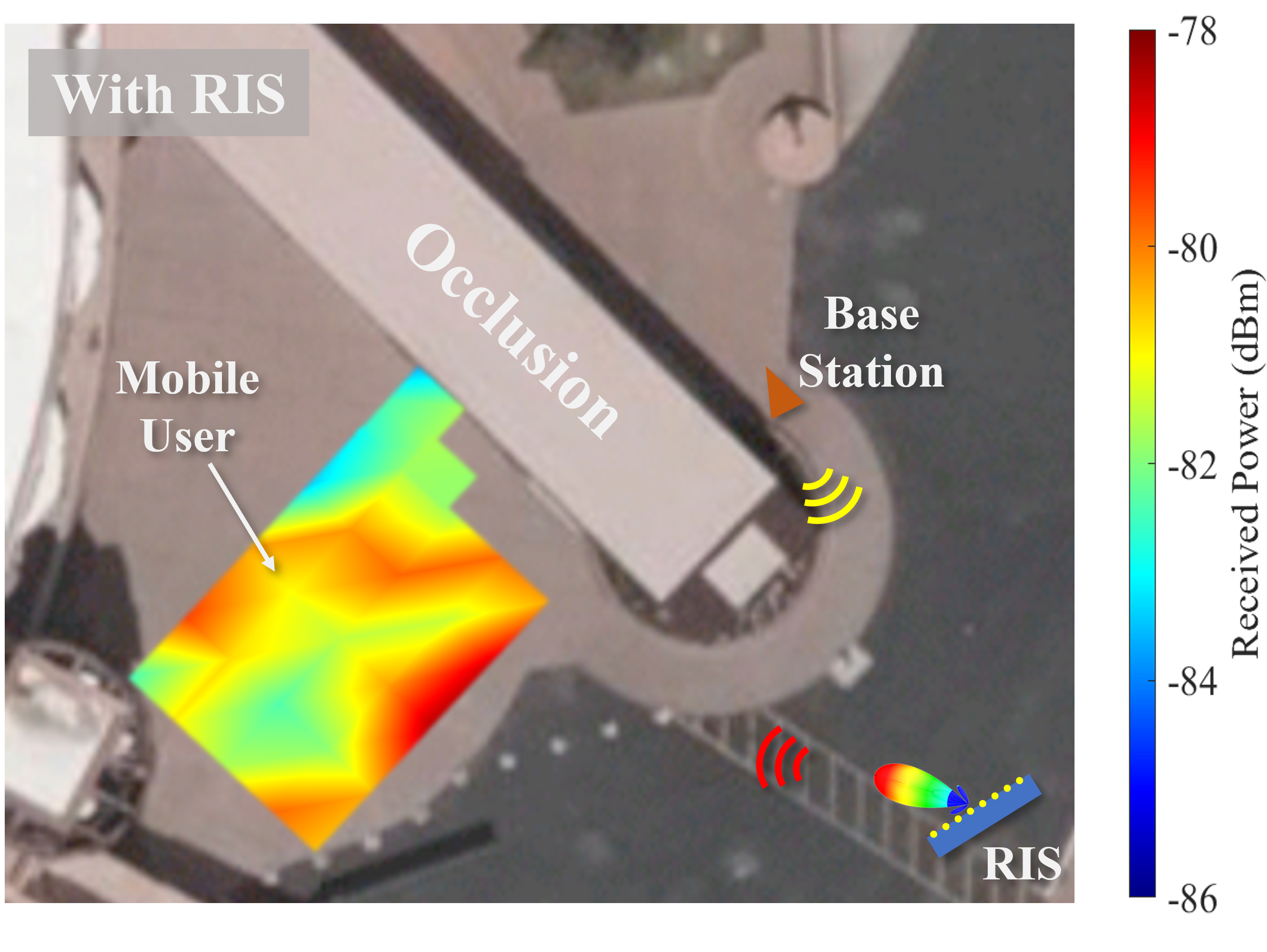}}
        \label{fig:with RIS}
    \end{subfigure}
\caption{This figure illustrates the field test setup with strong occlusion between the base station and the mobile user. Figure (a) shows the coverage map without the RIS while figure (b) shows the coverage map with the RIS. The field measurements indicate an average SNR gain of $\sim 6$dB (max 8 dB) when the RIS is deployed.}
\label{fig:Coverage}
\end{figure}

In this section, we present the results of the signal coverage measurements. 
The goal is to evaluate the improvement of the received signal power for a mobile user that employs an omnidirectional antenna. This experiment was done using the measurement setup 2 (ASU Gammage Memorial Auditorium), described in \sref{subsec:setup}. In the absence of RIS, the received signal on a 28-point grid is low due to the occlusion (signal blockage) from the large wall, as shown in the coverage map in \figref{fig:Coverage}(a). To evaluate the potential of the RIS in expanding the signal coverage, we place the RIS at a strategic place such that the RIS is in LoS of both the base station and the occluded area (in front of the venue entrance). As such, the RIS intercepts the signal from the base station and redirects it to the mobile user. In this scenario, the path from the base station to the RIS to the UE ranges between 30-40 meters depending on the grid position. As the user moves to each grid position, the RIS scans the codebook beams horizontally to find the optimum signal reception. The second coverage map in \figref{fig:Coverage}(b) shows the improved received signal level using the RIS beam scanning. \textbf{The SNR improvement along the occluded/blocked region is up to 8 dB with an average of 6dB. We note here that the SNR improvement is a function of the RIS size, thus a reasonably 10 times larger aperture (1,600 elements) would result in 28 dB SNR improvement for the same coverage scenario.} The 20 dB improvement is due the fact that the received signal depends on the square of the RIS area, as in (1) and (3).

\section{Conclusion} 
In this work, we developed a proof-of-concept prototype for reconfigurable intelligent surfaces and evaluated their potential gains in real-world environments and practical settings. In particular, we designed and fabricated a sub-6GHz 160-element RIS prototype, which relies on  a planar  single layer reconfigurable reflectarray (no vias) capable of scanning in both azimuth and elevation planes.  Thanks to the simplicity of the geometry and the capability to operate with single-bit switches, the proposed design can be directly scaled to higher frequencies such as mmWave and THz systems using either RF PIN switches,  
transistor based switches (e.g. CMOS-based), or  tunable materials (e.g. graphene). For this RIS prototype, we characterized the beamforming capabilities for both the passive (no RF diodes) and active implementations which achieved a half-power-beamwidth (HPBW) of approximately 9 degrees and 16 degrees on the azimuth and elevation planes, respectively. Then, we integrated the RIS into a wireless communication system to accurately evaluate its beamforming and coverage gains in realistic communication scenarios. Our results indicated that the developed RIS system can provide around $20$dB SNR gain when both the transmitter/receiver use directional antennas and when they are at distances of $5$m and $10$m from the RIS. Further, when the average BS $\rightarrow$ RIS $\rightarrow$ UE distance is $35$m, with blocked LoS link, and when only one side uses a directional antenna while the other side adopts an omni-directional antenna, an SNR gain of 8 dB is achievable. This gain can be further increased by increasing the size of the RIS.  For example, a ten-fold increase in the area of the RIS to a moderate 1,600-element array will further increase the SNR by 20. 

{This work has shown that RISs is a promising technology in extending wireless coverage in scenarios where occlusions are strong, even if the RIS is in the far field of both base station and mobile user.} Additionally, RISs have the potential to operate in extremely low power levels which is important for i) sustainable wireless communications and ii) to enable deployments where access to power supply is limited or even non-existent. The current non-optimized prototype consumes less than 0.4 Watt (including the biasing circuitry). As an alternative to PIN diodes, transistor-based switches \cite{venkatesh_high-speed_2020} are promising devices for extremely low power consumption RISs. The low leakage current of field-effect-transistor (FET) switches may consume less than 4 $u$Watt of DC power for an RIS with 10,000 switches. Finally, beamforming performance can be improved in future designs by eliminating the grating lobes that appear in the opposite direction to the main beam. This improvement can be achieved by adding a fixed, random phase delay on each unit cells and mitigate the quantization errors stemming from low-bit sampling \cite{kashyap_mitigating_2020,yin_single-beam_2020}.

\section{Acknowledgements}
We would like to thank Rogers Inc. Chandler, AZ for providing the substrates for the implementation of all prototypes presented in this work. Many thanks to Craig Birtcher for his help in the radiation pattern measurements in the Compact Antenna Test Range.

\linespread{1.5}

\end{document}

%% file: input.tex
\usepackage{amsfonts}
\usepackage{times}
\usepackage{graphicx}
\usepackage{latexsym}
\usepackage{dsfont}
\usepackage{amssymb}
\usepackage{amsmath}
\usepackage{cite}
\usepackage{verbatim}
\usepackage{subfigure}

\newcommand{\figref}[1]{{Fig.}~\ref{#1}}

\def\bb0{{\mathbb{0}}}

\def\ba{{\mathbf{a}}}
\def\bb{{\mathbf{b}}}

\def\bh{{\mathbf{h}}}

\def\b0{{\mathbf{0}}}

\def\bbE{{\mathbb{E}}}

\def\sf0{{\mathsf{0}}}

